\documentclass[preprint,12pt,authoryear]{elsarticle}
\pdfoutput=1
\makeatletter
\def\ps@pprintTitle{%
	\let\@oddhead\@empty
	\let\@evenhead\@empty
	\def\@oddfoot{}%
	\let\@evenfoot\@oddfoot}
\makeatother

\usepackage[dvipsnames,x11names]{xcolor}
\usepackage[colorlinks=true,linkcolor=blue, citecolor=blue, urlcolor=blue]{hyperref} 

\makeatletter
\AtBeginDocument{\def\@citecolor{DodgerBlue3}}
\makeatother

\makeatletter
\AtBeginDocument{\def\@citeauthorcolor{DodgerBlue3}}
\makeatother

\makeatletter
\AtBeginDocument{\def\@linkcolor{DodgerBlue3}}
\makeatother

\makeatletter
\AtBeginDocument{\def\@urlcolor{DodgerBlue3}}
\makeatother

\usepackage{natbib}
\usepackage{graphicx}
\usepackage{subfig}

\makeatletter
\def\maxwidth{\ifdim\Gin@nat@width>\linewidth\linewidth
	\else\Gin@nat@width\fi}
\def\maxheight{\ifdim\Gin@nat@height>\textheight\textheight
	\else\Gin@nat@height\fi}
\makeatother

\usepackage[lowercase]{theoremref}

\newtheorem{definition}{Definition}[section]

\usepackage{bbding}

\usepackage{algorithm}
\usepackage{algpseudocode}

\setkeys{Gin}{width=\maxwidth,height=\maxheight,keepaspectratio}
\usepackage{setspace}
\usepackage{calc}
\usepackage{graphicx}
\usepackage{subfig}
\usepackage{caption}
\usepackage{float}
\usepackage{colortbl}
\usepackage{longtable,booktabs}

\usepackage{color}

\usepackage{fancyvrb}

\usepackage{appendix}

\usepackage{amssymb,amsmath}
\DeclareMathOperator*{\argmax}{arg\,max}
\DeclareMathOperator*{\argmin}{arg\,min}

\usepackage{ifxetex,ifluatex}
\ifnum 0\ifxetex 1\fi\ifluatex 1\fi=0 
\usepackage{libertine}
\usepackage[scaled=0.95]{zi4}
\usepackage[libertine]{newtxmath}
\else 
\ifxetex
\usepackage{mathspec}
\usepackage{xltxtra,xunicode}
\else
\usepackage{fontspec}
\fi
\defaultfontfeatures{Mapping=tex-text,Scale=MatchLowercase}

\fi

\IfFileExists{upquote.sty}{\usepackage{upquote}}{}
\IfFileExists{microtype.sty}{%
	\usepackage{microtype}
	\UseMicrotypeSet[protrusion]{basicmath} 
}{}

\usepackage{epsfig}
\usepackage{amssymb}

\begin{document}
	
	\begin{frontmatter}

		\title{Applying Reinforcement Learning to Option Pricing and Hedging\tnoteref{t1}}
			\author{Zoran Stoiljkovic\corref{coraut} }	
			\cortext[coraut]{\textit{E-mail for correspondence:} \href{mailto:stoilj01@ads.uni-passau.de}{stoilj01@ads.uni-passau.de}, \href{mailto:zstoilj01@gmail.com}{zstoilj01@gmail.com}}
			
		\tnotetext[t1]{This is a slightly revised version of the thesis submitted in December 2022 in partial fulfillment of the requirements for the degree of \textit{Master of Science in Business Administration}. All errors and omissions are my own responsibility.}

		\begin{abstract}
			
This thesis provides an overview of the recent advances in reinforcement learning in pricing and hedging financial instruments, with a primary focus on a detailed explanation of the Q-Learning Black-Scholes approach, introduced by \citeauthor{halperin} (\citeyear{halperin}). This reinforcement learning approach bridges the traditional \citet{black1973} model with novel artificial intelligence algorithms, enabling option pricing and hedging in a completely model-free and data-driven way.  This paper also explores the algorithm's performance under different state variables and scenarios for a European put option. The results reveal that the model is an accurate estimator under different levels of volatility and hedging frequency. Moreover, this method exhibits robust performance across various levels of option's moneyness. Lastly, the algorithm incorporates proportional transaction costs, indicating diverse impacts on profit and loss, affected by different statistical properties of the state variables.

\

		\end{abstract}
		
		\begin{keyword}  QLBS \sep
		Dynamic Programming \sep  FQI \sep Offline Reinforcement Learning \sep Option Pricing \sep Hedging 
			
		\end{keyword}
		
	\end{frontmatter}
	
\section{Introduction}
Since \citet{turing} introduced the Turing test, discussing whether machines could think and imitate human behavior, computing powers have experienced enormous growth, such that nowadays machines can learn to imitate and perform specific tasks better than humans. Although the progress in \textit{Artificial Intelligence} (AI) is undeniable, it is important to maintain an awareness that the increased computational power of machines enables quicker but not necessarily accurate decisions \citep{russel2010}.

The cutting-edge achievements in \textit{Reinforcement Learning} (RL) resulted from successfully applying various advanced AI techniques to real problems. To a large extent, the progress made in this field is due to DeepMind's research. One of the first breakthroughs is the Deep Reinforcement Learning (DRL) model applied to Atari 2600 computer games, where an agent learned to play only from pixel inputs, outperforming all previous approaches and achieving superhuman performances \citep{atari2}. After a while, \citet{silver3} presented the computer program AlphaGo, which defeated the best professional human players in the board game Go using deep neural networks. Building upon these accomplishments, \citet{silver4} created the next AlphaGo Zero version. Following the tabula rasa principle and using a single neural network, AlphaGo Zero defeated its predecessor with the remarkable result of 100-0.

Although it may not seem as powerful as superhuman performance of AlphaGo, AI also makes its steps in finance. The interest in automated decision-making processes in finance is stronger than ever, as it can bring benefits to financial market participants. \citet{grobys5} found that the hedge funds which rely on AI and Machine Learning (ML) were superior in terms of the average profits compared to those with higher levels of human assistance. Drawing inspiration from \citet{silver4} and \citet{atari2}, \citet{kolm2019}  developed an analog version in finance, employing an RL model for option hedging, which achieved a lower cost than delta hedging.

There has been a continuously rising trend of publications in AI and ML in finance in recent years \citep{goodell6}. Despite the growing popularity of
AI and ML in finance research, there remains a need for critical evaluations regarding the practical implementations and benefits of constantly emerging advancements. Moreover, some contrary viewpoints about research in finance can be found. There are review studies that critique findings in financial economics due to replication failures, such as one of the often-cited studies by \citet{hou7}, which found that a significant portion of the results in finance fail to replicate. Conversely, \citet{chen2021open} demonstrated successful replication of nearly 100\% of the examined literature results, including those in \citeauthor{hou7}. The presence of such encouraging evidence provides a basis to maintain confidence in the credibility of a substantial body of existing findings in finance research.

In the first half of 2022, the value of outstanding over-the-counter (OTC) derivatives surged by 47\% compared to the previous year (\href{https://www.bis.org/publ/otc_hy2211.pdf}{BIS, end-June 2022}\nocite{OTC}). 
 With such substantial growth, it is anticipated that RL will find an expanding range of applications in derivatives pricing and hedging. This thesis primarily focuses on exploring the Q-Learner Black-Scholes (QLBS) model of \citeauthor{halperin} (\citeyear{halperin}) for pricing and hedging a European option. 
 
 \subsection{Outline and Objectives of the Thesis}
 
 \

 The rest of the paper is structured as follows:
 
 \begin{itemize}
 	
 	\item \textbf{Chapter \ref{chap_RL}} introduces readers to the \textit{Basic Principles of Reinforcement Learning}. This chapter also discusses \textit{Offline RL}, which has been underrepresented in the literature so far.
 	
 	\item \textbf{Chapter \ref{chap_3}} elaborates on the \textit{Black-Scholes-Merton} model.
 	
 	\item \textbf{Chapter \ref{ch_RL_options}} gives an overview of \textit{Reinforcement Learning in Pricing and Hedging Options}.  
 	\begin{itemize}
 		\item First, it presents a \textit{literature review}.
 		\item Further, I provide a thorough analysis of \textit{QLBS}, including the technical notes, which are given in \ref{app_technical}. Despite the seeming simplicity of the math involved, these notes have not been previously reported to the best of my knowledge at the time of this writing. 
 		Also, they may prove useful to students in Business Administration who may not have had the same level of exposure to mathematical concepts as those in natural and technical sciences.
 		
 		\item At the end of the chapter, the model-based and model-free QLBS are contrasted.
 	\end{itemize}
 	\item The \textit{Results}, including the \textit{simplified} exemplification of QLBS, are presented in \textbf{Chapter \ref{chap_5}}. At this stage can be defined  three basic research questions, which will be explored for three different states:
 	
\textit{RQ1)} What are the effects of different levels of \textit{volatility and hedging frequency} on QLBS pricing and hedging?
 	
 	\textit{RQ2)} How does the model perform at \textit{different moneyness} levels?

 \textit{RQ3)} What additional effects may arise from incorporating \textit{transaction costs}?
 	
 	\item \textbf{Chapter \ref{chap_6}} \textit{concludes} the paper, answering the research questions, and suggests possibilities for further research.
 \end{itemize}

 \section{Basic Principles of  Reinforcement Learning}\label{chap_RL} 
 
 One of the most widely accepted definitions of \textit{intelligence} explains it as "\textit{a very general mental capability that, among other things, involves the ability to reason, plan, solve problems, think abstractly, comprehend complex ideas, learn quickly and learn from experience}" \citep[p.13]{gottfredson1997}. While this definition is primarily associated with humans, planning, problem-solving and learning are also integral parts of AI. The term \textit{Artificial Intelligence} was coined by McCarthy in the 1950s, who later described AI as "\textit{the science and engineering of making intelligent machines, especially intelligent computer programs}" \citep [p.2]{AI_2004}.  Like humans, machines learn through a "\textit{trial-and-error}" procedure \citep{sutton2018}, where rewards and punishments can be part of learning \citep{turing}. In particular,  \textit{Reinforcement Learning} is the process of learning actions that maximize the cumulative reward by interacting with the environment.

 This chapter describes the basic principles of RL following the standard book of  \citet{sutton2018}, with slightly adjusted notations convenient for the RL problem setting in this paper.
 
 \subsection {Elements of Reinforcement Learning}\label{sect_2.1}
 
 \
 
 The main elements of RL are the \textit{goal-oriented} \textbf{agent} and \textbf{environment}. An agent can be considered as a learner that interacts with an environment and receives feedback based on this interaction. The agent can be imagined as a trader in finance, a player in computer games, or even the human brain, receiving dopamine as a reward for specific experiences \citep{lapan2020}. Alongside these two elements, there are also sub-elements: states, actions, and rewards.
 
 \
 
 \textbf{States, actions and rewards}
 
 \
 
 The agent's history up to time $t$ is defined as: 
 \begin{equation}
 	\mathcal{H}_t = O_1,  A_1, R_2, ... , O_{t-1}, A_{t-1}, R_t
 \end{equation}
where $O_t$ is the observation, $A_t$ is the action, and $R_t$ represents the reward at time-step $t$.  

The \textbf{state} $x$ is a function that maps the history to a state and can be formally represented as $X_t = f (\mathcal{H}_t)$ \citep{silver2015}. The tuple of states $x \in \mathcal{X}$ can be understood as a collection of features. When a stochastic process has the \textit{Markov property}, it means that the process satisfies the property of memorylessness, where the current state $X_t$ depends solely on the preceding state $X_{t-1}$, and not on the entire history of previous states. Consequently, the current state provides sufficient information for determining an optimal action. Formally, the Markov property can be represented as: 
 \begin{equation}
 	\mathbb{P}(X_t | X_{t-1}; X_{t-2}; \dots; X_0) = \mathbb{P}(X_t|X_{t-1})
 \end{equation}

\

 The next sub-element is the \textbf{action} $A_t$ that the agent takes at time $t$. Each action $a$ belongs to the set of all possible actions $\mathcal{A}$, and the set of actions available in state $x$ is $\mathcal{A}_x \subseteq \mathcal{A}$. 
 
 \
 
 Finally, the feedback that an agent receives from the environment, which is perceived as either bad or good, is called \textit{reinforcement} or reward \citep{russel2010}. In the context of RL, a \textbf{reward} may be understood as an overall goal of the learning process. The concept of reward originates from psychology, where experts make slight distinctions between the terms reward, reinforcement, and incentive \citep{wise1978}. Just as dopamine release in the human brain represents a positive signal for rewarding activities, the agent in an RL system receives a scalar reward $R_t$ from the environment at each time-step $t$, from the set of all possible rewards $\mathcal{R}$.

 \
 
 \textbf{State transition probabilities}
 
 \
 
 The dynamics from one state to another can be represented using a \textit{transition function} $p:\mathcal{X} \times  \mathcal{X}\rightarrow [0,1]$, or as:
 \begin{equation}
 	p (x'|x) = \mathbb{P} (x' = X_{t+1}|x = X_t), \quad \forall \{x, x'\} \in \mathcal{X}
 	\label{eq:state_trans}
 \end{equation}
 which implies that transition to state $x'$ is conditioned on the previous state $x$ and can be denoted as $\mathcal{P}_{x x'}$. Thus, the transition matrix $\mathbf{\mathcal{P}}$ with dimensions $|\mathcal{X}| \times |\mathcal{X}|$ can be represented as:

 \begin{equation}
	\mathbf{\mathcal{P}} = \begin{bmatrix}
		\mathcal{P}_{11} & \cdots & \mathcal{P}_{1n} \\
		\vdots & & \vdots \\
		\mathcal{P}_{n1} & \cdots & \mathcal{P}_{nn}
	\end{bmatrix}
\end{equation}
where each element $\mathcal{P}_{x x'}$ represents the probability of transitioning from state $x$ to state $x'$, and the sum of each row equals 1 \citep{silver2015}. 

\
 
\textbf{Markov Decision and Markov Reward Processes}
 
 \

 \begin{definition}[Markov Reward Process]
 A	Markov Reward Process (MRP) is a set of discrete-time steps and a tuple $ {\langle} \mathcal{X},\mathcal{P},\mathcal{R},\gamma {\rangle} $ with a finite set of states $\mathcal{X}$, state transition probabilities  $\mathcal{P}_{x x'}$, reward function $ \mathcal{R}_x = \mathbb{E} [R_{t+1}|x=X_t]$, and the discount factor $\gamma \in \{\mathbb{R}| 0 \leq \gamma \leq 1 \}$.
 \end{definition}
 In MRP, the reward $R_{t+1}$ depends only on the previous state $X_t$, while transition probabilities are defined in (\ref{eq:state_trans}).
 
 If we redefine the transition probability from equation (\ref{eq:state_trans}) to:
 \begin{equation}
 	p (x’, r|x,a) = \mathbb{P} (x’ = X_{t+1}, r=R_{t+1}|x = X_t, a= A_t), \quad \forall \{x, x'\} \in \mathcal{X}, r \in \mathcal{R}, a \in \mathcal{A}_x
 	\label{eq:state_trans_01}
 \end{equation}
 we get a function $p:\mathcal{X} \times \mathcal{R} \times \mathcal{X} \times \mathcal{A} \rightarrow [0,1]$, which depicts the dynamics of Markov Decision Processes, and can be briefly denoted as $\mathcal{P}^a_{xx'}$.
 \begin{definition}[Finite Markov Decision Process]
 A	finite	Markov Decision Process (MDP) is a set of discrete-time steps and a tuple $ {\langle} \mathcal{X}, \mathcal{A}, \mathcal{P}, \mathcal{R}, \gamma {\rangle} $ with a finite set of states $\mathcal{X}$, actions $\mathcal{A}$, transition probabilities $\mathcal{P}^a_{xx'} $, reward function $\mathcal{R}_{x,a} = \mathbb{E} [R_{t+1}|x=X_t, a=A_t]$, and the discount factor $\gamma \in \{\mathbb{R}| 0 \leq \gamma \leq 1 \}$.
 \end{definition}
 Compared to MRP, in MDP both the reward function and transition probabilities are extended for actions. 
 
 RL framework is formalized as a sequential decision-making problem at discrete-time points under uncertainty in a \textit{finite} MDP, where $|\mathcal{X}| < \infty$ and $|\mathcal{A}|<\infty$ \citep{russel2010}. In this process, an agent learns how to make \textit{good} actions $A_t \in \mathcal{A}$ by interacting with an environment.\footnote{At each time-step $t= 0,1,2,…,T $.} The agent is transitioning deterministically from the current state $X_t $ to the next state $X_{t+1}$, obtaining an appropriate reward $R_{t+1}$ for its action. By receiving the reward $R_{t+1}$ as the result of its previous action $A_t$, the agent observes the new state $X_{t+1}$.

 Any approach appropriate for solving MDP is considered to be an RL method.
 
 \
 
 \subsubsection{Model-Free vs. Model-Based Learning}\label{sec_2.1.2}
 
 \
 
 Since the model of the world can be known or unknown, in RL we can distinguish two categories: 
 \begin {itemize}
 \item 	\textit{Model-free} learning 
 \item 	\textit{Model-based} learning
 \end {itemize}
 
It is worth noting that in RL literature, the term \textit{model} does not relate to any statistical model used for learning. In RL, \textit{model-free} refers to situations when transition probabilities and a reward function are unknown, while the opposite is true for \textit{model-based}.

 \textit{Dynamic Programming} \citep{bellman1957} is a well-known technique which belongs to model-based learning. According to his autobiography, Richard Bellman first used the term \textit{dynamic programming} while working at RAND.\footnote{US company, based on research, established to advise United States Armed Forces.} To avoid using research-related words due to "\textit{pathological fear}" of his superior Secretary of Defense, he came up with an idea how to express "\textit{multistage decision processes}" \citep[p.159]{bellman1984eye}. Dynamic programming is a problem-solving technique that involves breaking down a problem into multiple sub-problems and solving each sub-problem only once if it arises multiple times. However, one of the limitations of dynamic programming is that it requires knowledge of the system dynamics, which are not always available in real-world applications.
 
 While model-based methods can predict the next state and reward and are thus suitable for planning, model-free RL may be understood as trial-and-error learning, applied when the environment dynamics are unknown. Fitted-Q-Iteration (FQI) and Q-learning are examples of model-free RL and will be covered in Section \ref{section_q_learn}.

\
 
 \subsubsection{Total Reward}\label{sec_2.1.3}
 
\

The overall goal of an agent is to maximize the total (cumulative) reward $(TR)$:\footnote{In literature often denoted as the return $G_t$.}
 \begin{equation}
 	{TR_t} = \gamma^0 R_{t+1} + \gamma^1 R_{t+2} + ... + \gamma^{T-1} R_{T} 
 	\label{eq_reward1}
 \end{equation}
 The discount factor $\gamma$ gives an opportunity to balance between present and future expectations. It can range from $0$ to $1$, depending on our goals. When choosing $\gamma$ to be 1, we sacrifice immediate higher rewards to achieve more promising ones over the long run. On the other hand, selecting $\gamma = 0$ indicates a sole emphasis on immediate reward. Choosing the values of $\gamma$ within the range of 0 to 1 allows us to trade off between these two extremes \citep{russel2010}. As discussed in \citeauthor{tsitsiklis2002} (\citeyear{tsitsiklis2002}), an alternative for discounted reward defined in equation (\ref{eq_reward1}) is the \textit{average reward}, which is obtained when $\gamma$ is close to 1.
 
 \
 
 \subsubsection{Value Functions and Policy}\label{sec_2.1.4}
 
 \
 
The goal of an agent in RL is to maximize the total reward by finding an optimal policy $\pi^\star$.
 \begin{definition}[Policy] A policy $\pi$ represents the agent's behavior at a certain time-step $t$. A \textbf{deterministic policy} always delivers the same action for a particular state and can be represented as $\pi:\mathcal{X} \rightarrow \mathcal{A}_x$, or $\pi(X_t)=A_t$. In contrast, a \textbf{stochastic policy} can be represented as $\pi (a|x) = \mathbb{P} (a=A_t | x=X_t)$ and may be understood as a probability distribution over $a \in \mathcal{A}_x$, $\forall x\in\mathcal{X}$, and in this case, different actions can be chosen in the same state. 
 \end{definition}
 In MDP, the policy $\pi$ depends only on the current state due to the Markov property. Finding an optimal policy $\pi^\star$ is a key step in RL and therefore is essential to define the functions which represent the conditional expectations of the total cumulative reward $\mathbb{E}(TR_t| \cdot)$:
 \begin{itemize}
 	\item \textbf{State-value function} under policy $\pi$: $V^\pi (x)$
 	\item \textbf{Action-value function (Q-function)} under policy $\pi$: $Q^\pi (x,a)$
 \end{itemize}
 \begin{definition}[State-value function] The state-value function $V^\pi(x)$ represents the expected total reward if the agent starts from the state $x=X_t$ and then follows the policy $\pi$. It can be recursively expressed as:\footnote{ Note that $\gamma^0 = 1$ and $\gamma^1 = \gamma$.}
 	\begin{equation}
 		\label{value}
 		\begin{split}
 			V^\pi (x) 		&= \mathbb{E}^\pi (TR_t|x=X_t)	\\		
 			&= \mathbb{E}^\pi (\gamma^0 R_{t+1} + \gamma^1 R_{t+2} + … +\gamma^{T-1}R_T|x=X_t)  \\
 			&= \mathbb{E}^\pi (R_{t+1} + \gamma TR_{t+1}|x=X_t)  \\ 
 		\end{split}
 	\end{equation}
 	for all $x \in \mathcal{X}$.
 \end{definition}
 \begin{definition}[Optimal policy] \label{def_optimal_policy} An optimal policy $\pi^\star$ is the policy that maximizes the expected total reward, $\pi^\star (x) = \max_{\pi} V^\pi (x)$, for all $x \in \mathcal{X}$. Consequently, this implies that  $V^{\pi^{\star}}(x) \geq V^\pi(x)$, $\forall x\in\mathcal{X}, \forall \pi \neq \pi^\star$.
 \end{definition}
 
 It is noteworthy that one policy $\pi$ is considered better than another policy $\pi'$ only if $V^\pi (x)$ is higher than $V^{\pi'} (x)$ for all states $x \in \mathcal{X}$. To determine the optimal policy $\pi^\star$, it is needed to compute the optimal value functions.

 \begin{definition}[Action-value function] The action-value function $Q^\pi(x,a)$ represents the expected total reward starting from the state $x=X_t$, taking action $a=A_t$, and following the policy $\pi$ afterward. It can be expressed formally as:
 	\begin{equation}
 		\label{Action_value1}
 		\begin{split}
 			Q^\pi (x,a) 		&= \mathbb{E}^\pi (TR_t|x=X_t, a=A_t)	\\		
 			&= \mathbb{E}^\pi (\gamma^0 R_{t+1} + \gamma^1 R_{t+2} + \cdots +\gamma^{T-1}R_T|x=X_t, a=A_t)  \\
 			&= \mathbb{E}^\pi ( R_{t+1} + \gamma TR_{t+1}|x=X_t, a=A_t).  \\
 		\end{split}
 	\end{equation}
 \end{definition}
 Similarly as defined in (\ref{def_optimal_policy}), the optimal policy $\pi^\star$ maximizes the action-value function $Q^\pi(x,a)$: 
 \begin{equation}
 	\label{optimal_policy}
 	\pi^\star (x) = \max_{\pi} Q^\pi (x,a), \quad \quad  \forall x \in \mathcal{X},  \forall a \in \mathcal{A}_x
 \end{equation}

 \subsection{Bellman Equations}\label{sec_2.2}
 
 \
 
 To find the optimal policy $\pi^\star$, we utilize well-established Bellman equations.

 \begin{definition}[Bellman expectation equations] The Bellman expectation equation decomposes the value functions into the conditional expectation of immediate reward and discounted value function at the next time-step. More specifically, using a recursive definition, the Bellman expectation equation for the action-value function can be represented as follows:
 	\begin{equation} %
\resizebox{1.1\textwidth}{!}{%
 		\label{bellman}
 	$	\begin{split}
 			Q^\pi (x,a) & =  \mathbb{E}^\pi \left[  TR_{t} | x= X_t, a=A_t \right]   \\			
 			&= \mathbb{E}^\pi \left[  R_{t+1} + \gamma TR_{t+1} | x= X_t, a=A_t \right]   \quad\quad \quad\quad \text{$\triangleright$ \footnotesize {Linearity of expectation}} \\
 			&= \mathbb{E}^\pi \left[  R_{t+1}|x=X_t, a=A_t\right] + \gamma \mathbb{E}^\pi [\mathbb{E}^\pi [TR_{t+1}|x' = X_{t+1}, a'=A_{t+1}]]  \quad \text{$\triangleright$ \footnotesize {Law of iterated expectations}}\\
 			&= \mathbb{E}^\pi \left[  R_{t+1}|x=X_t, a=A_t \right] + \gamma \mathbb{E}^\pi \left[ Q^\pi (x'=X_{t+1},a'=A_{t+1}) \right] \quad \text{$\triangleright$ \footnotesize {Linearity of expectation}}\\
 			&= \mathbb{E}^\pi \left[ R_{t+1} + \gamma Q^\pi(x'=X_{t+1},a'=A_{t+1})|x=X_t, a=A_t \right] \\
 			&= \mathcal{R}_x^a + \gamma \sum_{x' \in \mathcal{X}} \mathcal{P}_{xx'}^a \sum_{a' \in \mathcal{A}} \pi (a'|x') Q^\pi (x',a')\\
 		\end{split} $%
 	}
 	\end{equation}
 	Analogously, the Bellman expectation equation for the state-value function can be expressed as:	
 	\begin{equation}
 		\begin{split}
 			V^\pi(x) &= \mathbb{E}^\pi\left[ TR_t| x=X_t\right] \\
 			&=  \mathbb{E}^\pi\left[ R_{t+1} + \gamma TR_{t+1}| x=X_t\right] \\
 			& \vdots \\
 			&=  \mathbb{E}^\pi\left[ R_{t+1} + \gamma V^\pi (x'=X_{t+1})| x=X_t\right] \\
 			&= \sum_{a \in \mathcal{A}} \pi(a|x) \left[ \mathcal{R}^a_x + \gamma \sum_{x' \in \mathcal{X}}\mathcal{P}_{xx'}^a V^\pi (x') \right] \\
 		\end{split}
 	\end{equation}
 \end{definition}

\
 
 The obtained action-value function (or state-value function) does not represent the optimal Q-value that maximizes the expected total reward, but instead, it represents the value of a  particular state for a given action. On the other hand, Bellman optimality equations output the optimal state and action values under the optimal policy $\pi^\star$. The "\textit{Principle of Optimality}" defined by Richard Bellman states that: 
 \begin{quote}\textit{An optimal policy has the property that whatever the initial state and initial decision are, the remaining decisions must constitute an optimal policy with regard to the state resulting from the first decision} \citep[p.83]{bellman1957}. \end{quote}

\
 
 The Bellman optimality equation for the action-value function $Q^\star(x,a)$, which yields the maximum conditional expected total reward, can be represented as follows: 
 \begin{equation}
 	\begin{split}
 		Q^\star (x,a)  &= \mathbb{E} [ R_{t+1} + \gamma \max_{a'\in\mathcal{A}} Q^\star(x'=X_{t+1}, a'= A_{t+1})|x=X_t, a=A_t] \\
 		& = \mathcal{R}_x^a + \gamma \sum_{x' \in \mathcal{X}} \mathcal{P}_{xx'}^a \max_{a' \in \mathcal{A}} Q^\star (x',a')\\
 	\end{split}
 \end{equation}
 and the state-value function as:
 \begin{equation}
 	\begin{split}
 		V^\star (x)  &= \max_{a \in \mathcal{A}_x} Q^{\pi^\star} (x,a)\\
 		&= \max_{a \in \mathcal{A}_x} \mathbb{E}^{\pi^\star} \left[ R_{t+1} + \gamma TR_{t+1} | x=X_t, a= A_t\right] \\
 		&= \max_{a \in \mathcal{A}_x} \mathbb{E} \left[ R_{t+1} + \gamma V^\star (x'=X_{t+1}) | x=X_t, a=A_t \right] \\
 		&=\max_{a \in \mathcal{A}_x} \mathcal{R}_x^a + \gamma \sum_{x' \in \mathcal{X}} \mathcal{P}^a_{xx'} V^\star(x') \\
 	\end{split}
 \end{equation}

\
 
Finally, we arrive at the expressions for the optimal action- and state-value functions:
 \begin{equation}
 	\begin{split}
 		&Q^\star (x,a) = \max_\pi Q^\pi (x,a),  \quad \quad \forall x \in \mathcal{X}, \forall a \in \mathcal{A}_x \\
 		&V^\star (x) = \max_\pi V^\pi (x), \quad \quad \quad \quad \forall x \in \mathcal{X} \\
 	\end{split}
 \end{equation}

 \
 \subsection{Offline Reinforcement Learning}\label{sec_2.3}
 
 \
 
 This section addresses the existing research gap between two basic types of learning: \textit{online} and \textit{offline} (batch mode) RL. These two different approaches of learning can be intuitively understood as direct and indirect learning. While the agent in an online setting learns the optimal policy in real-time by interacting directly with an environment, the agent in the batch mode is expected to determine the best policy "\textit{indirectly}" without actively exploring the environment.\footnote{ In offline RL, an agent is not confronted with the exploration-exploitation dilemma, as in the online setting.} In offline RL, the agent is provided with a fixed dataset of completed interactions with an environment $\mathcal D_t^k=\left\{\left(X_t^k,A_t^k,R_t^k,X_{t+1}^k\right)\right\}_{t=0}^{T-1}$ for $k \in \{1,...,K\}$, including a history of states, actions, and rewards for each time-step and each $k$ simulation path \citep{dixon2020}.
 
In practice, it has been observed that the benefits of both approaches could be mutually used, resulting in the emergence of a new in-between approach, referred to as the "\textit{growing batch}" approach by \citet{offline1}. 
 The growing batch approach differs from batch learning because it incorporates exploration\footnote{This form of exploration can be understood as an agent not completely relying on the obtained information set, but also seeking additional experiences to enhance its learning and improve its policy-making.} as part of the learning process, aiming to extend the sample experience and improve the policy over time. 
 As pointed out by \citeauthor{levine2020} (\citeyear{levine2020}), it is important to distinguish between the terms "\textit{batch reinforcement learning}" and "\textit{batch}", since the latter is often used in ML to refer to the learning algorithm that uses a batch of data in the iterative learning process. While \citeauthor{levine2020} emphasize the potential of offline RL algorithms in utilizing pre-collected data, they also discuss the challenges posed by distributional shifts. According to \citeauthor{silver2021reward} (\citeyear{silver2021reward}), offline learning is particularly useful in addressing problems that have already been solved using similar data sets. Conversely,  \citeauthor{silver2021reward} argue that online learning allows an agent to address problems as they arise, leading to continuous knowledge improvement. Nonetheless, one significant advantage of offline RL is that it circumvents the challenges associated with exploration in the real world, which can often be impractical, time-consuming, dangerous, or costly \citep{levine2020}.

\ 

\begin{figure}[H]
 	\centering	\includegraphics[width=15cm,height=5cm,keepaspectratio]{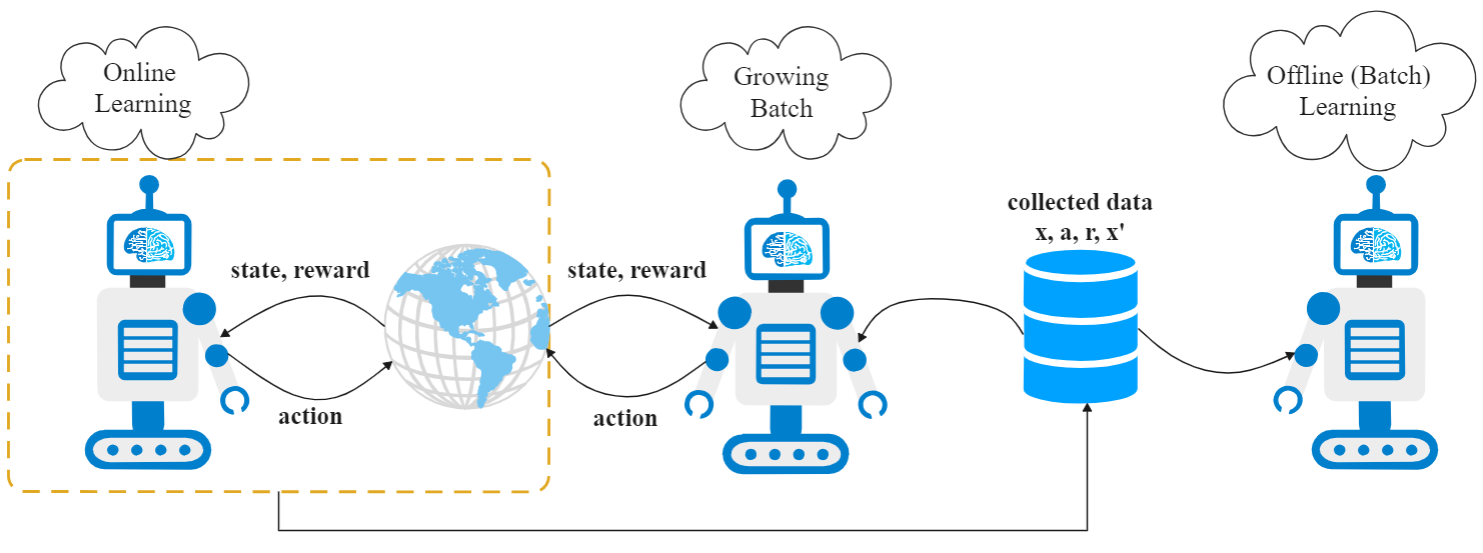}
 	\caption{Online, Growing Batch, and Offline RL}
 	\label{offline}
 \end{figure}

 Figure \ref{offline} provides a visual representation of the differences between online, growing batch, and batch mode learning discussed above.

 \newpage
 \subsection{Q-Learning and Fitted-Q-Iteration}\label{section_q_learn}
 
 \
 
 One of the solutions to non-linear Bellman optimality equations is the well-known \textit{Q-learning} algorithm \citep{watkins1989}. Q-learning belongs to the \textit{Temporal-Difference} \citep{sutton1988learning} learning family, where it is not required that the dynamics of the environment are known, but instead, an algorithm learns from experience and thus is a model-free RL method. In his PhD Thesis, \citeauthor{watkins1989} (\citeyear{watkins1989}) describes iterative action-value learning as the process that directly delivers the optimal action-value function $Q^\star(x,a)$.\footnote{The action-value function $Q(x,a)$ converges to the optimal action-value function $Q^\star(x,a)$ with probability 1 \citep{watkins1992q}.} The update in Q-learning is defined as:
 
 \begin{equation}
 	Q(x,a) \leftarrow Q(x,a) + \alpha \left[  r  + \gamma \max_{a'} Q(x', a') - Q(x,a) \right]
 	\label{q_learning}
 \end{equation}
 
 with learning rate $\alpha \in [0,1]$, which controls the weight of the update. While by setting $\alpha$ to zero, the agent would not take any new experience, and therefore, the action-value function would not be updated, the opposite is true when $\alpha$ equals 1.
 The second term in equation (\ref{q_learning}), $r  + \gamma  \max_{a'} Q(x', a') - Q(x,a)$, is known as the Temporal-Difference error,\footnote{ Often denoted as $\delta$ in literature.} that the algorithm attempts to minimize. One limitation of Q-learning is the \textit{maximization bias}, which arises because the algorithm consistently selects the action with the highest Q-value, potentially leading to an overestimation of the true optimal action values. To address this bias, Double-Q-Learning \citep{hasselt2010} splits the $max$ operator into two parts and employs two Q-values for the update process.
 
 The key distinction between Q-learning and FQI \citep{fqi01} is that the former updates the Q-value online, while the latter does it in an offline manner, using the collected experience from the sample set $\mathcal{D}$ \citep{riedmiller2005}. FQI can be useful for both discrete and continuous state spaces $\mathcal{X}$ \citep{dixon2020}. Additionally, FQI has the edge over Q-learning due to the possibility of being used with non-parametric function approximation \citep{fqi01}. The general representation of FQI may be understood as regression \citep{dixon2020}:
 \begin{equation}
 	r + \gamma \max_{a'}Q(x',a') = Q(x,a) + \epsilon_t
 \end{equation}
 where $\epsilon_t$ is the random noise with $\mathbb{E}(\epsilon_t) = 0$. One possibility is to represent $Q(x,a)$ as an expansion over basis functions and then calculate the parameters. Other possibilities include employing regression trees \citep{fqi01} or neural networks \citep{riedmiller2005} as function approximators. Another distinction of FQI is that it does not require processes to be Markovian \citep {murphy2005}.

 The algorithm pseudocodes of two simplified versions of both Q-learning and FQI are shown below:\footnote{ The representation of the FQI algorithm is adjusted for the implementation in this thesis, employing the backward update of the Q-function.}
 
 \begin{minipage}{0.44\textwidth}
 	\begin{algorithm}[H]
 		\centering
 		\caption{Q-learning}\label{algorithm_q_learning}
 		\footnotesize
 		\begin{algorithmic} 
 			\State \text{Input: learning rate $\alpha$, discount factor $\gamma$}
 			\State \text{Initialize $Q(x,a)$, $\forall x \in \mathcal{X}, a \in \mathcal{A}_x$} 
 			\For{\textbf{each} episode}
 			\State $\text{Initialize $x$}$
 			\Repeat{} 
 			\State \text{Choose action $a$ using policy $\pi$}
 			\State \text{Take action $a \in \mathcal{A}_x$, observe $r, x'$}
 			\State $Q(x,a) \leftarrow Q(x,a)  + $ 
 			\State {$\quad \quad \quad \quad \alpha     \: \left(r + \gamma \max_{a'} Q(x',a') - Q(x,a)\right)$}
 			\State {$x \leftarrow x'$}
 			\Until $\text{x is terminal}$
 			\EndFor
 		\end{algorithmic}
 	\end{algorithm}
 \end{minipage}
 \hfill
 \begin{minipage}{0.44\textwidth}
 	\begin{algorithm}[H]
 		\centering
 		\caption{FQI}\label{algorithm_fqi}
 		\footnotesize
 		\begin{algorithmic}
 			\State \text{Input: discount factor $\gamma$, $\mathcal{D} = \{(X^k_t, A^k_t, R^k_t, X^k_{t+1})\}$,  }
 			\State \text{$ \quad \quad$  for $t \in \{0,...,T\}$ and $k \in \{1,...,K\}$}
 			\State \text{Initialize $Q_T(x,a)$, $\forall x \in \mathcal{X}, a \in \mathcal{A}_x$} 
 			\For{$t=T-1, ... , 0$}					
 			\State $y:	r + \gamma \max_{a'}Q(x',a')$
 			\State $x: Q(x,a)$
 			\State regress $y$ on $x$
 			\State {$Q(x,a) \leftarrow Q(x',a')$}
 			\EndFor		
 		\end{algorithmic}
 	\end{algorithm}
 \end{minipage}

 \

 As can be seen from the algorithms above, the difference lies in the input parameters and the Q-value update. While Q-learning updates the Q-function by using one observation per update and thus has a slower convergence, FQI updates Q-values simultaneously by utilizing the cross-sectional data over all Monte Carlo paths \citep{dixon2020}.

 Overall, in Algorithm \ref{algorithm_q_learning}, the agent interacts with the environment over multiple episodes. The agent selects an action based on a policy (e.g., epsilon-greedy), receives a reward, and updates the Q-function. On the other hand, Algorithm \ref{algorithm_fqi} follows a batch learning approach. It takes a dataset $\mathcal{D}$ of observed experiences, including $X_t$, $A_t$, $R_t$, and $X_{t+1}$ for each time step. FQI initializes the Q-function, which is updated by regressing the target value $y$ on the current Q-function values.
 
 \
 
 \section{Cornerstones of Option Pricing and Hedging}\label{chap_3}

\begin{figure}[H]
 	\centering
 \includegraphics[width=14cm,height=14cm,keepaspectratio]{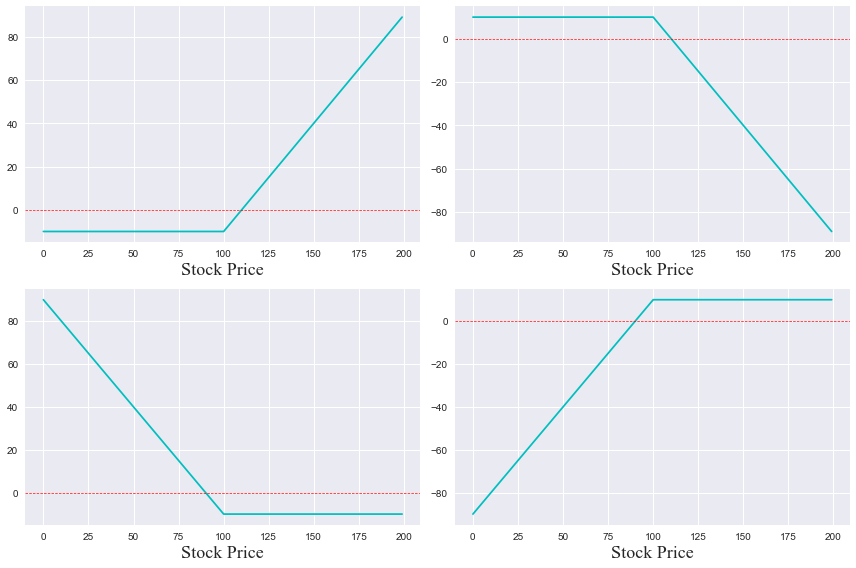}
 	\caption{Net profit from buying (left) and selling (right) an option with a strike price of 100\$} 
 	\label{fig_buy_sell}
 \end{figure}

 \begin{definition}[European Option]A European put (call) option gives the buyer the right to sell (buy) an underlying asset $S_t$ at a predefined strike price $Z$ at maturity (i.e. expiration date) $T$. 
 \end{definition}
 There are two types of participants in options markets:
 \begin{itemize}
 	\item Option buyer
 	\item Option seller (\textit{writer}) 
 \end{itemize}
 The \textit{buyer} has the \textit{right} to exercise the option, but not an obligation. On the other hand, the \textit{seller} receives an option premium,\footnote{Due to an asymmetric distribution of risks and chances, the buyer of an option pays a premium to the option seller.} but may need to fulfill the obligation if the buyer exercises the option, being therefore potentially exposed to losses \citep{HullBook}. The differences between these positions are shown in Figure \ref{fig_buy_sell}. 

 The Nobel prize-winning work of \citeauthor {black1973} (\citeyear{black1973}) and \citeauthor{merton1973} (\citeyear{merton1973}) (BSM) in option pricing and hedging serves as a key milestone for numerous models in modern quantitative finance up to today. Although subsequent approaches aimed to address the limitations of the BSM, such as  \citeauthor{leland}'s (\citeyear{leland}) incorporation of proportional transaction costs in discrete hedging, these models still build upon the BSM as the foundation for further expansions, refinements, and comparisons. This enduring reliance on the BSM underscores its exceptional significance as an achievement in the field of derivative pricing. Among the notable generalizations of the BSM is the discrete binomial-tree model proposed by \citeauthor{cox1979} (\citeyear{cox1979}). As $\Delta t \rightarrow 0$, the option price derived from the binomial tree model converges to the BSM price. 
 
 \subsection{Black-Scholes-Merton }\label{BS_01}
 
 \

 In their seminal paper, \citeauthor{black1973} (\citeyear{black1973}) introduced a model that continuously replicates an option using the hedge portfolio $\Pi$ consisting of the underlying stock and option. Later, some authors reinterpreted the hedge portfolio $\Pi$ using a bank account or bond instead of options (\citeauthor{bjork}, \citeyear{bjork}; \citeauthor{cochrane2005}, \citeyear{cochrane2005}). The hedge portfolio can be formulated as follows:
 \begin{equation}
 	\Pi_t = a_t S_t + b_t B_t
 	\label{eq_portfolio1}
 \end{equation}
 where $S_t$ represents an underlying asset at time $t$, $B_t$ is the risk-free zero bond, while $a_t$ and $b_t$ are positions in the stock and bond, respectively.\footnote{For further simplicity, a zero bond with face value of 1 USD at maturity $T$ is assumed.} The objective of dynamic replication is to achieve a perfect match between an option and its replicating (hedge) portfolio.

Following \citeauthor{bjork} (\citeyear{bjork}), the gain process $G_t$ of the portfolio $\Pi_t$ can be represented as:

 \begin{equation}
 	\label{gain}
 	G_t = \int_{0}^{t} a_u dS_u +  \int_{0}^{t} b_u dB_u
 \end{equation}
 where $dS_u$ and $dB_u$ denote the infinitesimal changes in the stock and bond price, respectively. Further, we can assume that the portfolio $\Pi_t$ is\textit{ self-financing}, which means that there are no further money inflows or outflows after the contract inception. The infinitesimal change of the portfolio $\Pi_t$ is then given by:
 \begin{equation}
 	\label {por}
 	\begin{split}
 		&d\Pi_t = d(a_t S_t + b_t B_t) = a_t dS_t + b_t dB_t \\
 		& \Pi_t - \Pi_0 = \int_{0}^{t} d(a_u S_u + b_u B_u ) \\
 		& \Pi_t = \Pi_0 + G_t \\
 	\end{split}
 \end{equation}
 From equation (\ref{por}), we have that $d\Pi_t = dG_t$, indicating the self-financing property of the constructed hedge portfolio $\Pi_t$. 
 
 The stock price follows a \textit{Geometric Brownian Motion} (GBM). The dynamics of the stock price $dS_t$ and the risk-free bond $dB_t$ are given by:
 
 \begin{equation}
 	\label{dynamics}
 	\begin{split}
 		&dS_t = \mu S_t dt + \sigma S_t dW_t\\
 		&dB_t = r B_t dt
 	\end{split}
 \end{equation}
 where $r$ is the risk-free interest rate, $\mu$ is the mean rate of stock return, $\sigma$ is the stock volatility, and $W_t$ is a standard Wiener process.

 \

Due to the \textit{law of one price}, the payoff of an option and hedge portfolio $\Pi_t$ must have an equal value,  ensuring that there are no opportunities for risk-free profit, in line with the \textit{no free lunch} principle. For further insights into the concept of \textit{no arbitrage}, please refer to \ref{appendix_BSM}.

\

 By substituting equation (\ref{dynamics}) into (\ref{por}), we can express the infinitesimal change of the portfolio $\Pi_t$ as follows:
 \begin{equation}
 	\begin{split}
 		d\Pi_t &= a_t dS_t + b_t d B_t = a_t\mu S_tdt+a_t\sigma S_tdW_t +  b_t r B_t dt \\
 		& = \left( a_t \mu S_t + b_t r B_t \right) dt + a_t \sigma S_t dW_t
 	\end{split}
 \end{equation} 
 
 As the option price depends on the underlying asset $S_t$, we can formally express the option as $C_t = F(S_t, t)$. By applying It\^{o}'s Lemma to the function $F (S_t,t)$, we obtain:\footnote{$dF_t(S_t,t) = F_s dS_t + F_t dt + \frac{1}{2} F_{ss}b(\cdot)^2 dt$, where $F_s$ is the first partial derivative, $F_{ss}$ is the second partial derivative of $F$ with respect to $S_t$, and $F_t$ is the partial derivative of $F$ with respect to $t$.}
 
 \begin{equation}
 	\label{derivative}
 	\begin{split}
 		dC_t &= \frac{\partial C_t}{\partial S_t} dS_t + \frac{\partial C_t}{\partial t} dt + \frac{1}{2} \frac{\partial^2 C_t}{\partial S_t^2} \sigma^2 S_t^2 dt \\
 		& = \frac{\partial C_t}{\partial S_t} (\mu S_t dt + \sigma S_t dW_t) + \frac{\partial C_t}{\partial t} dt + \frac{1}{2} \frac{\partial^2 C_t}{\partial S_t^2} \sigma^2 S_t^2 dt \quad\quad \text{$\triangleright$ By substituting (\ref{dynamics}) for $dS_t$}\\
 		& = \left(\frac{\partial C_t}{\partial S_t}\mu S_t +  \frac{\partial C_t}{\partial t} + \frac{1}{2} \frac{\partial^2 C_t}{\partial S_t^2} \sigma^2 S_t^2    \right) dt + \frac{\partial C_t} {\partial S_t} \sigma S_t dW_t \\
 	\end{split}
 \end{equation}
Since the value of the option $C_t$ must change by the same amount as the replicating portfolio $\Pi_t$, or $dC_t \overset{!}{=} d\Pi_t$, we have:
 \begin{equation}
 	\begin{split}
 		\left(\frac{\partial C_t}{\partial S_t}\mu S_t +  \frac{\partial C_t}{\partial t} + \frac{1}{2} \frac{\partial^2 C_t}{\partial S_t^2} \sigma^2 S_t^2    \right) dt + \frac{\partial C_t} {\partial S_t} \sigma S_t dW_t = \left( a_t \mu S_t + b_t r B_t \right) dt + a_t \sigma S_t dW_t
 		\label{BS2}
 	\end{split}
 \end{equation}
 Now, by equating $a_t = \frac{\partial C_t}{\partial S_t}$, we get:
 \begin{equation}
 	\begin{split}
 		\frac{\partial C_t}{\partial t} + \frac{1}{2} \frac{\partial^2C_t}{\partial S_t^2} \sigma^2 S_t^2 &= b_t r B_t \\
 		& = b_t r \left( \frac{C_t - a_tS_t}{b_t} \right) \quad\quad \text{$\triangleright$ Rewrite (\ref{eq_portfolio1}) as $B_t = \frac{C_t - a_t S_t}{b_t}$ }\\
 		& = r C_t - r a_t S_t \\
 		& = r C_t - r \frac{\partial C_t}{\partial S_t} S_t \\
 	\end{split}
 	\label{eq_bs3}
 \end{equation}
 After further rearranging of terms in (\ref{eq_bs3}), we obtain the well-known \citeauthor{black1973} partial differential equation:
 \begin{equation}
 	\frac{\partial C_t}{\partial t} + \frac{1}{2} \frac{\partial^2C_t}{\partial S_t^2} \sigma^2 S_t^2 - r C_t + r \frac{\partial C_t}{\partial S_t} S_t = 0
 	\label{BS1}
 \end{equation}
 with the terminal condition $C_T = max (Z-S_T; 0)$ for a European put option. The solution of equation (\ref{BS1}) for the put option $p$, following the representation from \citeauthor{HullBook} (\citeyear{HullBook}), is:
 \begin{equation}
 	p^{\left(BS\right)}=Ze^{-rT}\mathcal N\left(-d_2\right)-S_0\mathcal N\left(-d_1\right)
 \end{equation}
 with coefficients $d_1$ and $d_2$:
 \begin{equation}
 	\begin{split}		
 		&d_1= \frac{ln (\frac{S_0}{Z}) + (r+\frac{\sigma^2}{2})T}{\sigma \sqrt{T}},\\
 		&d_2 = d_1 - \sigma\sqrt{T}\\
 	\end{split}
 \end{equation}
We can interpret  $ \mathcal N(-d_2)$ as the probability for a European put option to be exercised. On the other hand, $ \mathcal N(-d_1)$ can be understood as a position in shares \citep{HullBook}.\footnote{ Have in mind that $\mathcal N(-d) = 1 - \mathcal N(d)$. }
 
\begin{figure}[h!]
 	\centering 	\includegraphics[width=10cm,height=5cm,keepaspectratio]{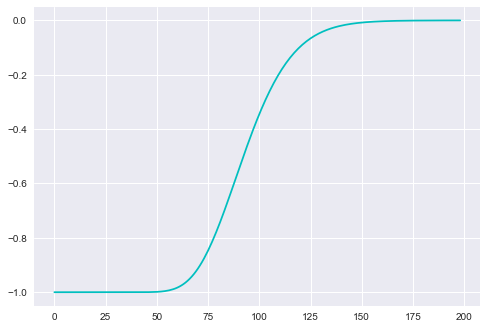}
 	\caption{Hedge ratio for a European put }
 	\label{hedge_ratio_BS}
 \end{figure}

 The hedge ratio $a_t = \frac{\partial C_t}{\partial S_t}$ is commonly referred to as delta ($\Delta$) in the literature.\footnote{Delta represents the sensitivity of the option price to changes in the underlying asset price.  A positive $\Delta$ indicates that the option price increases as the underlying asset price rises, while a negative delta indicates an inverse relationship.}  Figure \ref{hedge_ratio_BS} depicts the hedge ratio generated by the BSM model for a put option with a strike price Z=100. The negative hedge ratio implies that a short position in the underlying asset is necessary to hedge the put option's risk. Conversely, the hedge ratio for call options is positive, which indicates the need for a long position in the underlying asset for effective hedging.

\

The assumptions of the \citeauthor{black1973} model, such as continuous re-hedging, no transaction costs, and constant volatility, are based on ideal market conditions. In this light, it is worth quoting the following: \begin{quote}\textit{All models are approximations. Assumptions ... are never exactly true. All models are wrong, but some models are useful. ...  So since all models are wrong, it is very important to know ... what models are likely to produce procedures that work in practice (where exact assumptions are never true)} \citep[pp. 61-63]{box2009}. \end{quote}
 Whereas under the BSM assumptions options trading would not provide any advantages, in a real-life scenario with existing risk exposure, options and other OTC derivatives are far from redundant instruments.\footnote{In practice, one of the employed valuation approaches is the "\textit{practitioner Black-Sholes model}", which aligns each option's volatility parameter with its implied volatility \citep{hull2017}. } The review of the Bank of International Settlements shows that the notional value of outstanding OTC derivatives, including options, reached approximately \$ 632 trillion in the first half of 2022.\footnote {See more: \href{https://www.bis.org/publ/otc_hy2211.pdf}{BIS, OTC derivatives statistics at end-June 2022}. \nocite{OTC}} For instance, in April 2022, the average daily turnover in foreign exchange markets amounted to \$ 7.5 trillion, exceeding the daily global GDP by 30 times.\footnote{ See more: \href{https://www.bis.org/publ/qtrpdf/r_qt2212.pdf}{BIS, Quarterly review; December 2022}. \nocite{OTC1}} 
 
 While adjusting portfolios too frequently may be particularly suitable for academic research purposes, it incurs substantial transaction costs when implemented in practice. Therefore, it is more convenient to perform pricing and hedging at discrete time intervals. One such approach is QLBS, for which the solution converges to the continuous BSM when $\Delta t \rightarrow 0$, as shown by \citeauthor{halperin} (\citeyear{halperin}).

 \

 \section{Reinforcement Learning in Pricing and Hedging Options }\label{ch_RL_options}
 
The application of ML and AI approaches in pricing and hedging financial instruments is increasingly gaining popularity among both academic researchers and industry practitioners. This heightened interest arises from the effective utilization of advanced techniques, which offer process automation and optimization, leading to potential cost savings and efficiency improvements. While these approaches have already been widely adopted in other fields, their application in the realm of quantitative finance marks the start of a new era.

This chapter provides more detailed information on the relevant literature. It also elaborates on the QLBS approach of \citet{halperin}.                    
 
 \
 
 \subsection{Literature Review}\label{sec_4.1}
 
 \

In literature, many modern RL approaches for pricing and hedging derivatives are benchmarked with conventional methods, such as delta hedging. While some state-of-the-art models employ traditional RL, there is also an increase in the utilization of deep learning techniques, jointly forming DRL methods.\footnote{Following this line of reasoning, a distinction can be made between "\textit{hedgers}" and "\textit{deep hedgers}", depending on whether deep neural networks are employed to approximate hedging strategies.}
 
 After the introduction of the QLBS model by \citeauthor{halperin} (\citeyear{halperin}), which connects the traditional BSM and RL, a few authors have further developed models for pricing and hedging contingent claims, and this trend is expected to persist in the future. To address market frictions, such as trading costs, subsequent approaches have aimed to integrate them into their models (\citeauthor{deep_hedge}, \citeyear{deep_hedge}; \citeauthor{kolm2019}, \citeyear{kolm2019} etc.). \citeauthor{halperin1} (\citeyear{halperin1}) further extends the RL framework of QLBS to Inverse Reinforcement Learning (IRL), which determines an optimal policy observing only states and actions without explicit knowledge of rewards. Additionally, QLBS is shown to be applicable to option portfolios.
 
 \citeauthor{deep_hedge} (\citeyear{deep_hedge}) employ deep neural networks to approximate hedging strategies.\footnote{This has led to the emergence of the term "\textit{deep hedging}", which is already increasingly recognized in this field.} Their model not only incorporates market frictions but also includes multiple hedging instruments. Also, the authors provide a practical implementation by hedging the at-the-money European call option on the S\&P500 index. Additionally, \citeauthor{buehler2019} (\citeyear{buehler2019}) implement deep hedging for a portfolio of barrier options.\footnote{Given that Knock-Out products comprise approximately 50\% of the total exchange turnover in Germany's financial market (see more: \href{https://www.derivateverband.de/DEU/Statistiken/Boersenumsaetze}{DDV, Statistiken-Börsenumsätze})\nocite{DDV}, the application of RL to pricing and hedging barrier options might be of particular interest to market participants.} 
In a linked study, \citeauthor{murray2022} (\citeyear{murray2022}) present an actor-critic deep hedging algorithm that dynamically adjusts risk-aversion levels.
 
 \citeauthor{kolm2019} (\citeyear{kolm2019}) approximate the sarsa target utilizing non-linear techniques. Furthermore,
 \citeauthor{du2020} (\citeyear{du2020}) apply cutting-edge DRL methods, including the Deep Q-Networks (DQN), DQN with Pop-Art, and Proximal Policy Optimization (PPO), for hedging the European call options. 
 
 \citeauthor{cao2021} (\citeyear{cao2021}) implement the Deep Deterministic Policy Gradient (DPG) method to estimate the hedging cost's standard deviation, comparing the hedging performance of the accounting and cash flow approach. Further, \citeauthor{cao2022gamma} (\citeyear{cao2022gamma}) develop hedging strategies using the Distributed Deep Deterministic Policy Gradient (D3PG) algorithm for gamma and vega hedging.

Table \ref{tab:literature} outlines the key differences among the models discussed above, including whether RL or DRL approaches are employed, the techniques utilized, and whether transaction costs are incorporated. The column \textit{volatility} indicates the simulated volatility of the underlying assets, \textit{hedged item} specifies the financial instruments being hedged, while \textit{hedge instrument} represents the specific financial instruments utilized for hedging purposes. Finally, the last column relates to the consideration of the whole \textit{portfolio as the hedged item}. Also, these models differ in numerous other aspects, such as in treating transaction costs as proportional or non-linear.

 \begin{table}[H]
 	\renewcommand{\arraystretch}{1.3}
 	\centering
 	\caption{Literature review }
 	\resizebox{\textwidth}{!}{%
 		\begin{tabular}{@{}cccccccc@{}}
 			\toprule
 			\textbf{Authors}                                                 & \textbf{RL/DRL} & \multicolumn{1}{c}{\textbf{Approach(es)}}                                        & \textbf{\begin{tabular}[c]{@{}c@{}}Transaction \\ costs\end{tabular}} & \textbf{Volatility}                                                    & \textbf{Hedged item}                                                                          & \textbf{\begin{tabular}[c]{@{}c@{}}Hedge \\ instrument\end{tabular}}  & \textbf{\begin{tabular}[c]{@{}c@{}}Portfolio as the\\ hedged item?\end{tabular}} \\ \midrule
 			\rowcolor[HTML]{ECF4FD}
 			\begin{tabular}[c]{@{}c@{}}\citeauthor{halperin} \\  (\citeyear{halperin}) \end{tabular}       & RL              & QLBS                                                                             & \XSolidBrush                                                                     & deterministic                                                          & at-the-money put                                                                              & \begin{tabular}[c]{@{}c@{}}underlying \\ asset\end{tabular}           & \XSolidBrush                                                                     \\
 			\begin{tabular}[c]{@{}c@{}}\citeauthor{halperin1}   \\ (\citeyear{halperin1}) \end{tabular}       & (I)RL           & \begin{tabular}[c]{@{}c@{}}IRL; \\ QLBS for portfolio \\ of options\end{tabular} & \XSolidBrush                                                                      & deterministic                                                          & \begin{tabular}[c]{@{}c@{}}at-the-money put (IRL);\\ portfolio of options (QLBS)\end{tabular} & \begin{tabular}[c]{@{}c@{}}underlying\\ asset\end{tabular}            & \Checkmark                                                                   \\
 			\rowcolor[HTML]{ECF4FD}
 			\begin{tabular}[c]{@{}c@{}}\citeauthor{deep_hedge} \\  (\citeyear{deep_hedge}) \end{tabular} & DRL             & Deep hedge                                                                       & \Checkmark                                                                     & stochastic                                                             & at-the-money call                                                                             & \begin{tabular}[c]{@{}c@{}}portfolio of \\ derivatives\end{tabular}   & \XSolidBrush                                                                     \\
 			\begin{tabular}[c]{@{}c@{}} \citeauthor{buehler2019} \\  (\citeyear{buehler2019}) \end{tabular} & DRL             & Deep hedge                                                                       & \Checkmark                                                                     & stochastic                                                             & portfolio of barrier options                                                                  & \begin{tabular}[c]{@{}c@{}}portfolio of \\ derivatives\end{tabular}   & \Checkmark                                                                    \\
 			\rowcolor[HTML]{ECF4FD}
 			\begin{tabular}[c]{@{}c@{}}\citeauthor{kolm2019} \\ (\citeyear{kolm2019}) \end{tabular}     & RL             & \begin{tabular}[c]{@{}c@{}}Non-linear regression \\ to sarsa targets \end{tabular}                                                        & \Checkmark                                                                     & deterministic  & at-the-money call                                                                             & \begin{tabular}[c]{@{}c@{}}underlying\\ asset\end{tabular}            & \XSolidBrush                                                                     \\			
 			\begin{tabular}[c]{@{}c@{}}\citeauthor{du2020}\\  (\citeyear{du2020}) \end{tabular}      & DRL             & \begin{tabular}[c]{@{}c@{}}DQN;   \\ DQN with Pop-Art;\\ PPO\end{tabular}         & \Checkmark                                                                     & deterministic                                                          & \begin{tabular}[c]{@{}c@{}}at-, in-, and out-of-\\ -the-money call options\end{tabular}        & \begin{tabular}[c]{@{}c@{}}underlying\\ asset\end{tabular}            & \XSolidBrush                                                                     \\ 
 			\rowcolor[HTML]{ECF4FD}
 			\begin{tabular}[c]{@{}c@{}}\citeauthor{cao2021}  \\ (\citeyear{cao2021}) \end{tabular}     & DRL             & Deep DPG                                                                   & \Checkmark                                                                     & \begin{tabular}[c]{@{}c@{}}deterministic \& \\ stochastic\end{tabular} & at-the-money call                                                                             & \begin{tabular}[c]{@{}c@{}}underlying\\ asset\end{tabular}            & \XSolidBrush                                                                     \\
 			\begin{tabular}[c]{@{}c@{}} \citeauthor{murray2022} \\ (\citeyear{murray2022}) \end{tabular}  & DRL             & \begin{tabular}[c]{@{}c@{}} Actor-critic \\ deep hedging\end{tabular}                                                                    & \Checkmark                                                                     & stochastic                                                             & \begin{tabular}[c]{@{}c@{}}portfolio of OTC \\ derivatives\end{tabular}                       & \begin{tabular}[c]{@{}c@{}}liquid hedging \\ instruments\end{tabular} & \Checkmark                                                                   \\
 			\rowcolor[HTML]{ECF4FD}
 			\begin{tabular}[c]{@{}c@{}}\citeauthor{cao2022gamma}  \\  (\citeyear{cao2022gamma})\end{tabular}     & DRL             & D3PG                                                                             & \Checkmark                                                                    & \begin{tabular}[c]{@{}c@{}}deterministic \& \\ stochastic\end{tabular} & call option                                                                                   & at-the-money call                                                     & \XSolidBrush                                                                     \\ \bottomrule
 		\end{tabular}%
 	}
 	\label{tab:literature}
 \end{table}

 \subsection{QLBS}\label{4.2}
 
\

 \textbf{Replicating portfolio and option pricing}
 
 \
 
 As already stated, the main focus of this thesis is on the QLBS model of \citet{halperin}. 
 
 Similar to the BSM, the starting point in QLBS is the replicating portfolio $\Pi_t$, constructed from an underlying stock $S_t$ and cash $B_t$ (instead of a zero bond). Thus, equation (\ref{eq_portfolio1}) can be redefined as:
 \begin{equation}
 	\Pi_t = a_tS_t + B_t
 	\label{eq01}
 \end{equation}
 where $a_t$ represents the hedge at time-step $t$. By applying the self-financing constraint, the replicating portfolio in discrete time can be expressed as:\footnote{To get equation (\ref{eq_pi}) from (\ref{eq01}), please refer to \ref{app_1}.}
 \begin{equation}
 	\Pi_t = \gamma [\Pi_{t+1} - a_t \Delta S_t]
 	\label{eq_pi}
 \end{equation}
 with the discount factor $\gamma = exp(-r\Delta t)$ and $\Delta S_t = S_{t+1} - exp(r\Delta t) S_t$. The risk-free interest rate is denoted as $r$, and the difference between two sequential time-steps as $\Delta t $. Equation (\ref{eq_pi}) is used to calculate the value of the replicating portfolio for $t \in \{0, 1, ... ,T-1\}$, while at the last time-step $T$, the portfolio $\Pi_T$ equals the option payoff.\footnote {The payoff of the put option is $max (Z-S_T;0)$, where $Z$ is a strike price.}
 
 The estimated fair option price $\hat{C}_t$ at time $t$ is then given by:
 \begin{equation}
 	\hat{C}_t = \mathbb{E}_t \left[ \Pi_t | \mathcal{F}_t \right]
 \end{equation}
 However, this price does not account for the option seller's risk exposure. To address this, the writer should incorporate a risk premium given by the discounted variance of the replicating portfolio, scaled by the risk-aversion parameter $\lambda$, as suggested by \citeauthor{halperin} (\citeyear{halperin}):
 \begin{equation}
 	\begin{split}
 		C_0 (S,a) &= \hat{C}_0 + \lambda \mathbb{E}_0  \left[ \sum_{t=0}^{T}  exp(-r t) Var [\Pi_t|\mathcal{F}_t] \middle| S_0=S, a_0=a \right]  \\
 		& = \mathbb{E}_0  \left[ \Pi_0 + \lambda \sum_{t=0}^{T}  exp(-r t) Var [\Pi_t|\mathcal{F}_t] \middle| S_0=S, a_0=a \right]\\
 	\end{split}
 	\label{eq_option_price}
 \end{equation}
 To ensure the competitive option price, the option seller should aim to minimize the option price $C_0 (S,a)$. 
 
 \

 \textbf{Optimal hedge $a_t^\star$}
 
 \
 
 As mentioned in Chapter \ref{BS_01}, continuous portfolio rebalancing is infeasible. Hence, the goal of the option seller is to minimize the potential losses caused by mishedged positions in discrete time.  \citet{grau2008} outlines several ways for calculating $a_t^\star$:

\begin{enumerate}
	\item\textit{\textbf{Global Hedging }represents the variance-risk minimization of portfolio $\Pi$ at $t=0$:
		\begin{equation}
			a_t^\star = \argmin_{a} \left[Var\left(\Pi_{0}|S_{0},t_0 \right)\right]
	\end{equation}}
	\item\textit{\textbf{Local Hedging} is a strategy that aims to minimize the variance between the replicating portfolio $\Pi_t$ and option value $C_t$:
		\begin{equation}
			\begin{split}
				a_t^\star = \argmin_{a} \left[Var\left(\Pi_{t+1} - C_{t+1}| S_t, t \right)\right]
			\end{split}
		\end{equation}
	}
	\item \textit{\textbf{Forward Global Hedging} does not consider previous information, but instead minimizes  variance as follows:
		\begin{equation}
			a_t^\star = \argmin_{a} \left[Var (\Pi_{t})| S_t, t\right]
			\label{fg_hedge}
		\end{equation}    
		which effectively reduces the next hedging errors. }
\end{enumerate}

 We will follow the third approach of minimization, as in \citeauthor{halperin} (\citeyear{halperin}). Given that, we can focus on equation (\ref{fg_hedge}) and rearrange it accordingly: 
 \begin{equation}
 	\begin{split}
 		a_t^\star&= \argmin_a Var [\Pi_t|\mathcal{F}_t]\\
 		&= \argmin_a Var  \left[ \gamma (\Pi_{t+1} - a_t \Delta S_t) |  \mathcal{F}_t \right] \quad\quad \text{$\triangleright$ By applying (\ref{eq_pi}) for $\Pi_t$}\\
 		&= \argmin_a \big[ \mathbb{E}_t \left[(\Pi_{t+1} - \mathbb{E}_t({\Pi}_{t+1}))^2\right] - \quad \quad\quad \quad \text{$\triangleright$ For the steps, see (\ref{eq_variance_portfolio})} \\
 		&  \quad\quad\quad\quad 2 a_t \mathbb{E}_t  \left[  (\Pi_{t+1} - \mathbb{E}_t (\Pi_{t+1}))  (\Delta S_t - \mathbb{E}_t (\Delta S_t))  \right]+ a_t^2 \mathbb{E}_t \left[ (\Delta S_t - \mathbb{E}_t (\Delta S_t) )^2 \right]   \big] \quad\quad \\
 	\end{split}
 	\label{eq_optimal_hedge}
 \end{equation}
 By setting the derivative of (\ref{eq_optimal_hedge}) to zero and rearranging it, we obtain:
 \begin{equation}
 	\begin{split}
 		a_t^\star &= \frac{\mathbb{E}_t \left[ (\Pi_{t+1} - \mathbb{E}_t (\Pi_{t+1})) (\Delta S_t - \mathbb{E}_t (\Delta S_t)) \right]}{\mathbb{E}_t \left[( \Delta S_t - \mathbb{E}_t(\Delta S_t))^2 \right]}\\
 		&= \frac{\mathbb{E}_t \left[ \hat{\Pi}_{t+1} \Delta \hat{S}_t \right]}{\mathbb{E}_t \left[ (\Delta \hat{S}_t)^2 \right]}\\
 		&= \frac{Cov(\Pi_{t+1}, \Delta S_t| \mathcal{F}_t)}{Var (\Delta S_t| \mathcal{F}_t)}
 	\end{split}
 	\label{eq_optimal_2}
 \end{equation}
 The optimal hedge $a_t^\star$ can be calculated based on the cross-sectional data from Monte Carlo simulation for each time-step $t$, starting from the time-step $T-1$. In a continuous setting, the expected values can be computed using expansions in basis functions.
 
\
 
 \textbf{State variables}
 
 \
 
 This paper explores how QLBS can learn by considering different statistical properties of three types of state variables:
 \begin{enumerate}
 	\item \textit{ The stock price dynamics defined in (\ref{dynamics}), which we repeated here for convenience:
 		\begin{equation*}
 			dS_t = \mu S_t dt + \sigma S_t dW_t
 		\end{equation*}
 		The GBM process follows a log-normal distribution, which is convenient for stock price simulation because it ensures non-negativity. Stock prices are non-stationary with a drift parameter $\mu$. }
 	\item \textit{We will also consider stock price returns. 
 		By applying It\^{o}'s Lemma to $F(S_t, t) = ln(S_t)$, we have:
 		\begin{equation}
 			\begin{split}
 				d ln(S_t) &= \frac{1}{S_t} dS_t + 0 + \frac{1}{2} (-\frac{1}{S_t^2}) \sigma^2 S_t^2 dt \\
 				&= -\frac{1}{2} \sigma^2 dt + \frac{1}{S_t} dS_t \quad\quad\quad\quad  \\ 
 				&= -\frac{1}{2} \sigma^2dt + \frac{1}{S_t} \left(  \mu S_t dt + \sigma S_t dW_t\right) \\
 				&=  -\frac{1}{2} \sigma^2dt + \frac{1}{S_t} \mu S_t dt + \frac{1}{S_t} \sigma S_t d W_t\\
 				&= \left(  \mu  -\frac{1}{2} \sigma^2   \right)dt + \sigma d W_t \\
 			\end{split} \label{eq_dln}\end{equation} Note that $ln(S_T) - ln(S_0) \sim \mathcal{N} \left( (\mu-\frac{1}{2} \sigma^2)T, \sigma^2 T \right)$, for $S_0 =1$.
 	}
 	\item \textit{	\citeauthor{halperin} (\citeyear{halperin}) uses the transformed variable $X_t$ to eliminate drift:
 		\begin{equation}
 			\begin{split}
 				dX_t &= -\left( \mu - \frac{1}{2} \sigma^2 \right) dt + d ln S_t \quad\quad\quad \text{ $\triangleright$ Apply (\ref{eq_dln}) for $dlnS_t$}\\
 				&= \sigma dW_t\\
 			\end{split}
 		\end{equation}	Given that the stock price $S_t$ follows GBM, the transformed variable $X_t$ is a martingale, so that $\mathbb{E}(dX_t) = 0$.}
 \end{enumerate}

 \begin{figure}[H]
 	\centering
 	\includegraphics[width= 14cm, height=6cm, keepaspectratio]{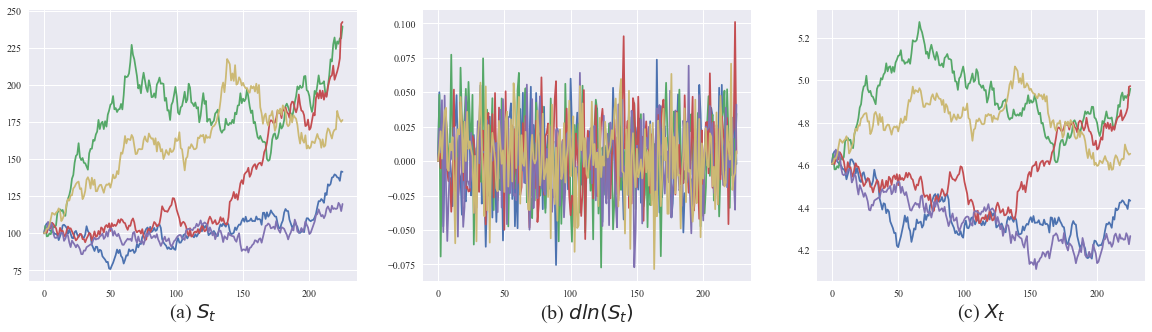}
 	\caption{State variables $S_t$, $dln(S_t)$, and $X_t$, using $K=5$ Monte Carlo paths}
 	\label{fig_states}
 \end{figure}
 
 In Figure \ref{fig_states}, it can be noticed that each of the states exhibits different properties. In the first case, when the state variable is $S_t$, stock prices demonstrate non-stationary behavior with drift, as dictated by the properties of GBM. The state variable in the second case $dln (S_t)$ represents stock returns, where $\mathbb{E} (dln(S_t)) \neq 0$. In the third case, the state variable $X_t$ is non-drifting and has the martingale property.\footnote{All three state variables are simulated in discrete time-steps using an Euler scheme, such that $dt = \Delta t$, and a Wiener process is $\Delta W_t = \sqrt{\Delta t} \varepsilon_t$, where $\varepsilon_t \sim \mathcal{N}(0,1)$.} 
 
 To be in line with \citeauthor{halperin} (\citeyear{halperin}), we denote states as $X_t$ in defining the RL elements of QLBS. However, the results encompass all three above-defined states, including  $S_t$ and $d ln(S_t)$. 
 
 \
 \
 \
 
 \textbf{Optimal action-value function and reward function in QLBS}
 
 \
 
As previously mentioned, the objective of the option seller is to minimize the asked option price given in equation (\ref{eq_option_price}). In the RL framework, the action-value function at time $t$, $Q_t(x,a)$, needs to be maximized. Hence, it can be expressed as:
 \begin{equation}
 	\argmax Q_t(x,a) = \argmin C_t (S, a)
 \end{equation}
 This implies that $Q_t = - C_t$. Thus, the action-value function under policy $\pi$ can be represented as:
 \begin{equation}
 	\resizebox{1\textwidth}{!}{%
$ 	\begin{split}
 		Q^\pi_t(x,a) &= \mathbb{E}_t \left[ - \Pi_t(X_t) - \lambda \sum_{t'=t}^{T} exp(-r (t'-t)) Var [\Pi_{t'}(X_{t'})| \mathcal{F}_{t'}] \middle| \mathcal{F}_t   \right] \quad\quad\text{$\triangleright$ Negative of  (\ref{eq_option_price}) } \\
 		&= \mathbb{E}_t \bigg[ - \Pi_t(X_t) - \lambda Var [\Pi_t (X_t)]  \quad \quad\quad  \qquad  \text{$\triangleright$ See \ref{append_action_value} for details}\\
 		& \quad\quad  - \lambda \sum_{t'=t+1}^T exp(-r (t'-t)) Var [\Pi_{t'}(X_{t'})|\mathcal{F}_{t'}] | \mathcal{F}_t \bigg] \\
 		&= \mathbb{E}_t \left[ -\Pi_t(X_t) -\lambda Var(\Pi_t(X_t)) + \gamma (Q_{t+1}^\pi (X_{t+1}, a_{t+1}) + \mathbb{E}_{t+1} [\Pi_{t+1} (X_{t+1})]) \right]  \\
 		&= \mathbb{E}_t \left[ R_t(X_t, a_t, X_{t+1}) + \gamma Q_{t+1}^\pi (X_{t+1},a_{t+1}) |x=X_t, a= a_t \right] 
 		\\
 		\label{qlbs_q_value}
 	\end{split} $%
}
 \end{equation}
 where $\mathbb{E}_t\left[ \space \; \cdot \; \space | \mathcal{F}_t\right] = \mathbb{E}_t\left[ \; \space \cdot \; \space | x=X_t, a=a_t \right] $. In this way, the Bellman expectation equation defined in equation (\ref{bellman}) has been recovered. 
 
 From equation (\ref{qlbs_q_value}), the reward can be expressed as:
 \begin{equation}
 	\label{reward}
 	\begin{split}
 		R_t\left(X_t,a_t,X_{t+1}\right)& =  \gamma \Pi_{t+1} (X_{t+1}) - \Pi_t(X_t) - \lambda Var \left[ \Pi_t\space|\space\mathcal F_t\right]  \\	
 		&= \gamma \Pi_{t+1} (X_{t+1}) - \gamma[\Pi_{t+1} - a_t \Delta S_t] - \lambda Var\left[\Pi_t\space|\space\mathcal F_t\right] \quad \text{$\triangleright$ By applying (\ref{eq_pi}) for $\Pi_t$ }\\
 		&= \gamma a_t\Delta S_t(X_t, X_{t+1})-\lambda Var\left[\Pi_t\space|\space\mathcal F_t\right] \\
 	\end{split}
 \end{equation}
 for $t= T-1, T-2, ..., 0$. The first term in equation (\ref{reward}) represents the return on the replicating portfolio, and the second term is the quadratic risk scaled by the risk-aversion parameter $\lambda$. Although it may not be immediately apparent that the reward is quadratic in actions $a_t$, this is shown in \ref{app_reward}, and it is the key feature that enables the analytic solution for $a_t$. 
 Due to the nature of the replicating portfolio $\Pi_T$ at expiry, where all stocks are converted into cash (i.e., $a_T = 0$), the reward at the last time-step is $R_T = -\lambda Var[\Pi_T]$.

The optimal action-value function is derived by combining equations (\ref{reward}) and (\ref{qlbs_q_value}): 
 
 \begin{equation}
 	\label{eq_q_value}
 	\begin{split}
 		Q_t^\star (X_t, a_t)  &= \gamma \mathbb{E}_{t} [ 	Q_{t+1}^\star(X_{t+1}, a_{t+1}^\star) + a_t \Delta S_t] - \lambda Var[\Pi_t (X_t)] \\
 		&  = \gamma \mathbb{E}_{t} [ 	Q_{t+1}^\star (X_{t+1}, a_{t+1}^\star) + a_t \Delta S_t] - \lambda \gamma^2 \mathbb{E}_t[\hat{\Pi}^2_{t+1} -2a_t\hat{\Pi}_{t+1} \Delta \hat{S}_t + a_t^2 (\Delta \hat{S}_t)^2 ] \\
 		& = \gamma \mathbb{E}_{t} [ 	Q_{t+1}^\star (X_{t+1}, a_{t+1}^\star) + a_t \Delta S_t] - \lambda \gamma^2 \mathbb{E}_t[ \left( \hat{\Pi}_{t+1} - a_t (\Delta \hat{S}_t)\right)^2 ] \\
 	\end{split}
 \end{equation}
 for $t \in \{T-1 ,\cdots, 0 \}$, while at $t=T$, $Q_T^\star(X_T,a_T=0) = -\Pi_T(X_T) - \lambda Var [\Pi_T(X_T)]$.
 
 \
 \
 
 \textbf{Optimal hedge in QLBS $a_t^\star$}
 
 \
 
 From equation (\ref{eq_q_value}), we can obtain the analytic solution for $a_t^\star (X_t)$ due to the quadratic term of $a_t$:\footnote{The derivation for the analytic solution, given by equation (\ref{a_star}), is shown in \ref{app_optimal_hedge}.}
 \begin{equation}
 	\label{a_star}
 	a_t^\star\left(X_t\right)=\frac{\mathbb{E}_{t} \left[  \Delta \hat{S}_{t}  \hat{\Pi}_{t+1} + \frac{1}{2 \gamma \lambda} \Delta S_{t} \right]}{
 		\mathbb{E}_{t} \left[ \left( \Delta \hat{S}_{t} \right)^2 \right]}
 \end{equation}
 In contrast to the global forward hedging \citep{grau2008} defined in equations (\ref{eq_optimal_hedge}) and (\ref{eq_optimal_2}), the optimal hedge in QLBS is extended by $\mathbb{E}_t(\Pi_t)$.
 
 Now, the analytic solution for the optimal hedge needs to be implemented in practice. Two possible approaches for implementation are dynamic programming and FQI. Recall that while the latter does not assume that transition probabilities and rewards are known, this assumption is necessary for dynamic programming. Within the QLBS framework, Monte Carlo simulation is employed to approximate the solution. 
 
 \
 
 \textbf{Model-based solution in QLBS}
 
 \
 
 In dynamic programming, the optimal hedge $a_t^\star$ and the action-value function $Q_t^\star$ are approximated using linear expansion over $N$ basis functions $\Phi_n\left(X_t^k\right)$, where $k$ denotes the $k$-{th} Monte Carlo simulation. Therefore, the optimal hedge $a_t^\star$ can be expressed as:
 \begin{equation}
 	\label{basis_01} 	a_t^\star\left(X_t\right)=\sum_n^N{\phi_{nt}\Phi_n\left(X_t\right)}\quad\quad
 \end{equation}
 with
 	$$
 	\mathbf{\phi}_t = \begin{bmatrix}
 		\phi_1 & \phi_2 & \cdots & \phi_N
 	\end{bmatrix} $$
 where $\mathbf{\phi}_t$ are the coefficients calculated by minimizing the negative of (\ref{eq_q_value}).\footnote{Note that maximizing $Q_t^\star (X_t,a_t)$ is equivalent to minimizing its negative. } The detailed steps can be found in \ref{app_coeff_phi}, where it becomes evident that the only difference between equations (\ref{a_star})  and (\ref{basis_01}) is the presence of basis functions in the latter.

 Similarly, the optimal Q-function can be represented as:
 \begin{equation}
 	\label{q_star}
 	Q_t^\star\left(X_t,a_t^\star\right)=\sum_n^N{\omega_{nt}\Phi_n\left(X_t\right)}
 \end{equation}
 where the parameters $\omega$ are calculated in a similar manner to the coefficients $\phi$. Note that $\omega_{nt}$ and $\phi_{nt}$ are time-dependent. The solution for determining $\omega_t$ is provided in \ref{app_optimal_omega}.

\
 
\begin{figure}[ht]
	\centering
	\subfloat[\centering order = 1]{\includegraphics[width=0.3\textwidth]{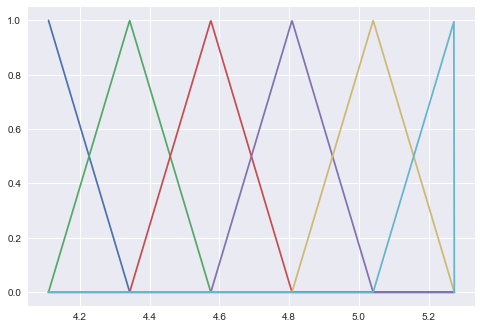}}\quad
	\subfloat[\centering order = 2]{\includegraphics[width=0.3\textwidth]{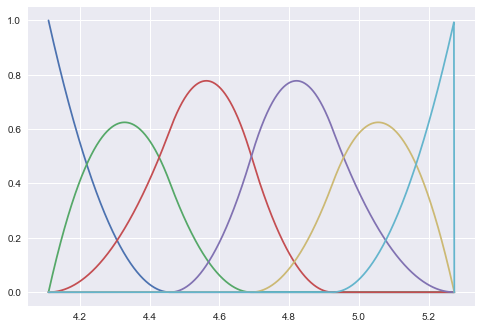}}\quad
	\subfloat[\centering order = 4]{\includegraphics[width=0.3\textwidth]{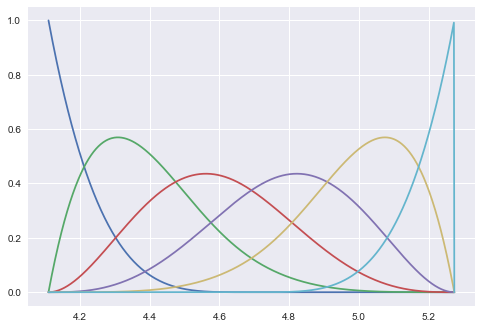}}
	\caption{B-splines ($N=6$) of different orders used for function approximation}
	\label{fig:b_splines}
\end{figure}

 As discussed by \citeauthor{grau2008} (\citeyear{grau2008}), the choice of basis functions can have a significant impact on function approximation. \citeauthor{grau2008} compared polynomial basis functions and splines and found that polynomial basis functions are suitable in up to three-dimensional problems. However, when considering a single dimension, splines exhibit lower errors than polynomials. \citet{halperin1} presents the results utilizing B-splines. 
 Figure \ref{fig:b_splines} provides an illustration of different orders of B-splines.

The algorithm pseudocode of the model-based QLBS is shown below:
\begin{algorithm}[H]
\caption{Model-based QLBS}\label{alg:DP}
\begin{algorithmic}		
	\State Simulate stock prices $S_{t}^k$ and calculate the state variable $X_t^k$ with $K$ Monte Carlo paths and $T$ time-steps 

	\State Determine the feature matrix $\left\{\Phi_n\left(X_t^k\right)\right\}_{n=1}^N$ with dimensions $T \times K \times N$, where $N$ represents the number of basis functions
	
	\State At $t=T$,  $\Pi_T = argmax (Z-S_{T},0) $; $\hat{\Pi}_{T} = \Pi_T - \mathbb{E}(\Pi_T)$;   $a_T = 0$; $R_T = -\lambda Var(\Pi_T)$;  $Q_T^{\star} = -\Pi_T - \lambda  Var(\Pi_T) $
	
	\For{$t$ from $t=T-1$ to $t=0$}
	
	\State Compute $a_t^{\star} (X_t)$ as in (\ref{basis_01})
	\State Compute $\Pi_t$ as in (\ref{eq_pi})
	
	\State Compute $R_t^{\star}$ as in (\ref{reward}) 
	
	\State Compute $Q_t^{\star} (X_t,a_t)$ as in (\ref{q_star})
	\EndFor
	
	\State  Calculate the QLBS option price at $t=0$ as $QLBS_{t=0} = - \frac{1}{K} \sum_{k=1}^{K} Q_{t=0}^{\star, k}$
\end{algorithmic}
\end{algorithm}

\textbf{ Model-free QLBS}

\

When transition probabilities and reward function are unknown, QLBS relies solely on data samples $\mathcal{D}_t^k=\left\{\left(X_t^k,a_t^k,R_t^k,X_{t+1}^k\right)\right\}_{t=0}^{T-1}$ for $k = 1,\ldots,K$, which can be either simulated or real data.  In FQI, the lack of the reward function and transition probabilities is compensated by richer input data, due to the availability of the action $a_t$ and reward $R_t$.

In FQI, the action-value function $Q_t^\star(X_t,a_t)$ is decomposed into the vector $\mathbf{A}_t$, matrix of coefficients $\mathbf{W}_t$, and vector of basis functions $\mathbf{\Phi}(X_t)$:
\begin{equation}
\mathbf{A}_t=\begin{bmatrix} 1\\a_t\\ \frac{1}{2}a_t^2\end{bmatrix}, \quad \mathbf{W}_t = \begin{bmatrix} W_{11}(t) & \cdots & W_{1N}(t) \\
	W_{21}(t) & \cdots & W_{2N}(t) \\
	W_{31}(t) & \cdots & W_{3N}(t) \\
\end{bmatrix} \quad \text{and} \quad \mathbf{\Phi}(X_t)=\begin{bmatrix} \Phi_1({X_t})\\\vdots\\ \Phi_N({X_t})  \end{bmatrix}
\end{equation}
The only unknown is the matrix $\mathbf{W}_t$, while both $\mathbf{A}_t$ and $\mathbf{\Phi}(X_t)$ can be easily determined from the model-based QLBS.

For FQI, we can use a set of basis functions as in the model-based QLBS. However, these basis functions depend on both the state $X_t$ and the action $a_t$. Specifically, the basis functions  $\mathbf{\Psi}(X_t,a_t)$ depend on the quadratic hedge $a_t$, and their dependence on the state $X_t$ is determined by the order of B-splines. The set of basis functions $\mathbf{\Psi} (X_t,a_t)$ can be obtained as:
\begin{equation} 
\begin{split}		
	& \vec{\mathbf{\Psi}}(X_t,a_t) = vec \left( \mathbf{A}_t \otimes \mathbf{\Phi}^T (X) \right)\\
\end{split}
\label{eq_basis_psi}
\end{equation}
where $\otimes$ denotes the outer product, and  $ \vec{\mathbf{\Psi}}(X_t,a_t)$ is the concatenated vector.

The optimal action-value function can be then represented as the following regression:
\begin{equation}
\begin{split}
	R_t (X_t, a_t, X_{t+1}) + \gamma \max_{a_{t+1} \in \mathcal{A}}   Q^\star_{t+1}(X_{t+1}, a_{t+1}) &= Q^\star_{t}(X_{t}, a_{t}) + \epsilon_t\\
	&= \mathbf{A}_t^T \mathbf{W}_t \mathbf{\Phi}(X)  + \epsilon_t\\
	&= \sum_{i=1}^{3} \sum_{n=1}^{N} \left( \mathbf{W}_t \odot (\mathbf{A}_t  \otimes \mathbf{\Phi}^T(X)) \right)_{in} + \epsilon_t \\
	& = \vec{\mathbf{W}}_t \vec{\mathbf{\Psi}} (X_t,a_t) + \epsilon_t  \quad\quad\\
\end{split}
\end{equation}
where the vector $\vec{\mathbf{W}}_t$ is the concatenated matrix $\mathbf{W}_t$, $\mathbf{E}(\epsilon_t)=0$, and $\odot$ denotes the element-wise product.\footnote{Note that \citeauthor{halperin} (\citeyear{halperin}) shows an additional solution for $Q_t^\star(X_t,a_t)$ to address potential overestimation bias. This alternative formulation is expressed as $Q_t^\star(X_t,a_t) =\mathbf{A}_t^T \mathbf{W}_t \mathbf{\Phi}(X) = \mathbf{A}_t^T \mathbf{U}_w(t,X_t)$, which can be applied once the coefficients $\mathbf{W}_t$ are computed. For a more comprehensive discussion, please refer to \citeauthor{halperin} (\citeyear{halperin}).} In a similar manner to the approach demonstrated in (\ref{app_eq_omega}), the time-dependent coefficients $\vec{\mathbf{W}}_t$ are determined through a least squares optimization process:
\begin{equation}
	\resizebox{1\textwidth}{!}{%
$
\vec{\mathbf{W}}_t = \left[ \sum_{k=1}^{K}{\Psi_n\left(X_t^k,a_t^k\right)\Psi_m\left(X_t^k,a_t^k\right)}\right]^{-1} \left[ \sum_{k=1}^{K}{\Psi_n\left(X_t^k,a_t^k\right)\left(R_t\left(X_t^k,a_t^k,X_{t+1}^k\right)+\gamma\max_{a_{t+1}\in\mathcal{A}}Q_{t+1}^\star\left(X_{t+1}^k,a_{t+1}\right)\right)}\right]
$%
}
\end{equation}
where $\left[\Psi_m\left(X_t^k,a_t^k\right)\right]^T=\Psi_n\left(X_t^k,a_t^k\right) $. The parameters $\mathbf{W}_t$ are computed backward starting  from $t=T-1$.

FQI is an off-policy algorithm, which means that it learns from suboptimal actions rather than strictly following a \textit{greedy} policy. In FQI, suboptimal actions are obtained by adding noise to the optimal actions $a_t^\star$.\footnote{The level of noise is determined  by the parameter $\eta \in [0,1]$.} Such additional noise does not present a challenge in the model-free QLBS. Moreover, FQI has the ability to determine the optimal action-value function $Q^\star(x,a)$ even with purely randomized actions $a_t$, as long as sufficient data is provided \citep{halperin}. As such, FQI offers a model-free solution to MDPs in a completely data-driven manner, which is a significant advantage due to no underlying assumptions.

\

\textbf{
Model-based vs. model-free QLBS}

\

Table \ref{tab:DP_vs_FQI_1} summarizes the key distinctions between the model-based and the model-free QLBS. 

\begin{table}[H]
\renewcommand{\arraystretch}{1.3}
\caption{Differences between the model-based and model-free QLBS}
\centering
\label{tab:DP_vs_FQI_1}
\resizebox{\textwidth}{!}{%
	\begin{tabular}{@{}ll@{}}
		\toprule
		\multicolumn{1}{c}{\textbf{Model-based QLBS}}                                                                                            & \multicolumn{1}{c}{\textbf{Model-free QLBS}}                                                                                                         \\ \midrule
		\rowcolor[HTML]{ECF4FD} 
		\begin{tabular}[c]{@{}l@{}}Expansion of the action-value function in the \\ state dependent set of basis functions $\Phi (X_t)$\end{tabular} & \begin{tabular}[c]{@{}l@{}}Expansion of the action-value function in the \\ action-state dependent set of basis functions $\Psi (X_t,a_t)$\end{tabular} \\
		\rowcolor[HTML]{FFFFFF} 
		Only the optimal hedge $a_t^\star$                                                                                                       & Does not use the optimal hedge                                                                                                                           \\
		\rowcolor[HTML]{ECF4FD}
		Compute actions and rewards                                                                                                           & Observe actions and rewards                                                                                                                         \\
		\rowcolor[HTML]{FFFFFF}
		Input is only the state variable $X_t$                                                                                                       & \begin{tabular}[c]{@{}l@{}}Input is the data set \\ $\mathcal{D}_t=\left\{X_t,a_t,R_t,X_{t+1}\right\}$\end{tabular}   \\
		\rowcolor[HTML]{ECF4FD}
		\begin{tabular}[c]{@{}l@{}}Outputs are the optimal action $a_t^\star$\\ and the optimal action-value function $Q_t^\star(X_t,a_t)$\end{tabular}  & \begin{tabular}[c]{@{}l@{}}Output is just the optimal action-value function \\ $Q_t^\star(X_t,a_t)$\end{tabular}                                              \\
		\begin{tabular}[c]{@{}l@{}}2N unknown parameters:\\ N for approximating $\phi_t$ and\\ N for approximating $\omega_t$\end{tabular}       & \begin{tabular}[c]{@{}l@{}}3N unknown parameters\\ for the parameter matrix $\mathbf{W}_t$\end{tabular}                                                  \\
		\rowcolor[HTML]{ECF4FD}
		\begin{tabular}[c]{@{}l@{}}For each time-step $t$ are given $1\times K$  observations\\ (provided only the state $X_t$)\end{tabular}                & \begin{tabular}[c]{@{}l@{}}For each time-step $t$ are given $3\times K$ observations\\ (for $X_t,a_t,R_t$)\end{tabular}                                     \\ \bottomrule
	\end{tabular}%
}
\end{table}

\

\section{Results}\label{chap_5}

\subsection{Simplification}

\

Inspired by the representation of \citeauthor{longstaff2001} (\citeyear{longstaff2001}) as well as \citeauthor{grau2008} (\citeyear{grau2008}), this section presents a simplified example of the model-based QLBS to show the computational steps.

Initially, we simulate the stock price $S_t$ with $S_0 = 100, \mu=0.05, \sigma=0.15, r=0.03$, and maturity of one year. For illustrative purposes, the number of Monte Carlo paths is set to $5$.

\begin{table}[H]
	\centering
	\caption{Simulated paths of stock prices}
	\begin{tabular}{@{}rrrrr@{}}
		\toprule
		\cellcolor[HTML]{FFFFFF}\textbf{} & \multicolumn{1}{c}{\cellcolor[HTML]{FFFFFF}\textbf{0}} & \multicolumn{1}{c}{\cellcolor[HTML]{FFFFFF}\textbf{1}} & \multicolumn{1}{c}{\cellcolor[HTML]{FFFFFF}\textbf{2}} & \multicolumn{1}{c}{\textbf{3}} \\ \midrule
		\rowcolor[HTML]{ECF4FD} 
		\textbf{1}                        & 100                                                    & 118.27                                                 & 124.43                                                 & 127.10                         \\
		\rowcolor[HTML]{FFFFFF} 
		\textbf{2}                        & 100                                                    & 86.20                                                  & 85.25                                                  & 83.75                          \\
		\rowcolor[HTML]{ECF4FD} 
		\textbf{3}                        & 100                                                    & 100.58                                                 & 96.50                                                  & 97.38                          \\
		\rowcolor[HTML]{FFFFFF} 
		\textbf{4}                        & 100                                                    & 97.20                                                  & 87.87                                                  & 96.10                          \\
		\rowcolor[HTML]{ECF4FD} 
		\textbf{5}                        & 100                                                    & 109.33                                                 & 128.43                                                 & 130.66             \\ \bottomrule    
	\end{tabular}
	\label{tab:my-table_GBM}
\end{table}

The strike price $Z$ is set to $100$, and the terminal payoff is $max (Z-S_T; 0)$. At the last time-step  $t=T=3$, the value of the replicating portfolio $\Pi_T (S_T)$ is equal to the terminal payoff, resulting in:

\begin{equation*}
\resizebox{\textwidth}{!}{%
$
	\mathbf{	\Pi_T} =\begin{bmatrix}
		0.00\\
		16.25\\
		2.62\\
		3.90\\
		0.00\\	
	\end{bmatrix},\quad
	\mathbf{	\hat{\Pi}_T = \Pi_T} - \overline{\Pi}_{T} =
	\begin{bmatrix}
		-4.55\\
		11.70\\
		-1.93\\
		-0.65\\
		-4.55\\	
	\end{bmatrix}\quad  \text{ and } \quad 
	\mathbf{Q_T^\star = - \Pi_T - }\lambda Var(\mathbf{\Pi_T}) = \begin{bmatrix}
		-0.04\\
		-16.29 \\
		-2.66\\
		-3.94\\
		-0.04\\
	\end{bmatrix} $ }
\end{equation*}
with the risk aversion $\lambda = 0.001$. Recall that at the final time-step $T$, the cash $B_T$ is equal to the terminal portfolio $\Pi_T$, as $a_T=0$. After computing all the elements of the final time-step $T$, we can proceed to $t=2$, obtaining:  

$$ 
\mathbf{\Delta S_2 = S_3} - exp(r\Delta t) \mathbf{S_2} =
\begin{bmatrix}
	1.42 \\
	-2.36 \\
	-0.09 \\
	7.35\\
	0.94\\
\end{bmatrix}, \quad
\mathbf{
	\Delta \hat{S}_2 = \Delta S_2} - \Delta \overline{S}_{2} = 
\begin{bmatrix}
	-0.03\\
	-3.81\\
	-1.54\\
	5.90\\
	-0.51 \\
\end{bmatrix} 
$$

\

\begin{figure}[H]
	\centering
	\includegraphics[width= 5cm, height=5cm, keepaspectratio]{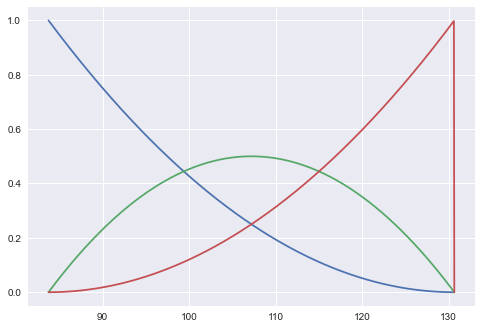}
	\caption{Three B-splines of the second order}
	\label{fig_basis}
\end{figure}

To approximate the hedge, we use the second-order B-splines with three basis functions, as depicted in Figure \ref{fig_basis}.

In a real numerical example, a higher number of basis functions is employed, but for the sake of simplicity, we limit it to three in this case. The feature matrix of B-splines $\mathbf{\Phi}_2 (S_2)$ is then constructed as:\footnote{Please note that here we use $S_t$ for the state variable.}

$$
\mathbf{\Phi_2} =\begin{bmatrix}
	
	0.02 & 0.23& 0.75\\
	0.94& 0.06& 0. \\
	0.53& 0.4 & 0.07 \\
	0.83& 0.16& 0.01\\
	0.  & 0.09& 0.91\\
\end{bmatrix} 
$$
Note that for each observation of $S_2$, the feature matrix $\mathbf{\Phi}_2$ consists of three values, corresponding to three basis functions employed for approximation. 

\

To calculate the optimal hedge, we need to determine the coefficients $\phi_2$, as shown in (\ref{a_06}). We can slightly adjust the coefficients as follows: 

\begin{equation*}
	\begin{split}
		\mathbf{	\phi_2 } &= \left[ \mathbf{\Phi_2^T} (\mathbf{\Phi_2 } \odot \left(\mathbf{\Delta \hat{S}_2}\right)^2)\right]^{-1} \big[ \mathbf{\Phi_2}^T\left[ \mathbf{\hat{\Pi}_3 \odot \Delta \hat{S}_2}\right] \big] \\
		& =  \begin{bmatrix}
			37.47 & 5.94 & 0.38 \\
			5.94 & 1.33 & 0.14 \\
			0.38 & 0.14 & 0.23 \\
		\end{bmatrix}^{-1} 
		\begin{bmatrix}
			-44.51 \\ -1.86 \\ 2.38 \\
		\end{bmatrix} 
		= 
		\begin{bmatrix}
			-3.05 \\ 11.37 \\ 8.2\\
		\end{bmatrix} \\
	\end{split} 
\end{equation*}

After obtaining the vector of coefficients $\phi_2$, we calculate the optimal hedge $a^\star_2$ using equation (\ref{basis_01}):

\begin{equation*}
	\mathbf{a^\star_2} = \mathbf{\Phi_2 \phi_2} = \begin{bmatrix}
		
		0.02 & 0.23& 0.75\\
		0.94& 0.06& 0. \\
		0.53& 0.4 & 0.07 \\
		0.83& 0.16& 0.01\\
		0.  & 0.09& 0.91\\
	\end{bmatrix} 
	\begin{bmatrix}
		-3.05 \\ 11.37 \\ 8.2 \\
	\end{bmatrix}  = 
	\begin{bmatrix}
		8.70 \\ -2.18 \\ 3.51 \\ -0.63 \\ 8.49\\
	\end{bmatrix}
\end{equation*}
Due to the limited number of simulated paths, some hedge values are non-negative and exhibit higher values. Once the optimal hedge ratio is calculated, we can determine the portfolio value and the reward:
$$
\mathbf{\Pi_2}= \mathbf{\gamma} \left[ \mathbf{\Pi_3} - a^\star_2 \odot \mathbf{\Delta S_2} \right] =
\begin{bmatrix}
	-12.24 \\
	10.98\\
	2.91\\
	8.45\\
	-7.90\\
\end{bmatrix} \quad \text{and} \quad
\mathbf{R_2 = \gamma \Pi_3 - \Pi_2 }- \lambda Var(\mathbf{\Pi_2})  = 
\begin{bmatrix}
	12.15 \\
	5.02\\
	-0.39\\
	-4.67\\
	7.81\\
\end{bmatrix}
$$
where the discount factor  $\gamma = exp (-r \Delta t) = 0.99$.

With all the necessary components, we can proceed to calculate the vector of coefficients $\omega_2$, as shown in equation (\ref{app_eq_omega}):
\begin{equation*}
	\begin{split}
		\mathbf{\omega_2} &= \mathbf{ \left[ \Phi_2^T \Phi_2 \right]^{-1} \big[\Phi_2^T \left( R_2 + \gamma Q_3 \right) \big]}\\
		& = \begin{bmatrix}
			1.85 & 0.41 & 0.06 \\
			0.42 & 0.25 & 0.28 \\
			0.06 & 0.28 & 1.40
		\end{bmatrix}^{-1}
		\begin{bmatrix}
			-18.91\\
			0.24 \\
			15.87\\
		\end{bmatrix} = 
		\begin{bmatrix}
			-12.85 \\ 10.71 \\ 9.74 \\
		\end{bmatrix}
	\end{split}
\end{equation*}
\
\

Finally, we can determine the optimal model-based Q-value for the time-step $t=2$:
\begin{equation*}
\mathbf{	Q_2^\star = \Phi_2 \omega_2} = \begin{bmatrix}
		
		0.02 & 0.23& 0.75\\
		0.94& 0.06& 0. \\
		0.53& 0.4 & 0.07 \\
		0.83& 0.16& 0.01\\
		0.  & 0.09& 0.91\\
	\end{bmatrix} 
	\begin{bmatrix}
		-12.85 \\ 10.71 \\ 9.74 \\
	\end{bmatrix} = \begin{bmatrix}
		9.51\\
		-11.41\\
		-1.84
		\\
		-8.85 \\
		9.83\\
	\end{bmatrix}
\end{equation*}

These steps outline the process for calculating all the elements of the model-based QLBS. Similarly, we obtain the Q-values and the optimal hedges for each path and time-step:
$$
\mathbf{a^\star} = \begin{bmatrix}
	-0.05 &	-2.86&	8.70&	0\\
	-0.05&	-4.02&	-2.18&	0\\
	-0.05&	-0.25&	3.51&	0\\		-0.05&	-0.63&	-0.63&	0\\
	-0.05&	-0.55&	8.49&	0\\
\end{bmatrix} 
\quad \text{and} \quad
\mathbf{Q^\star} =
\begin{bmatrix}
	-2.38 &	-4.10 &	9.51&	-0.04 \\
	-2.38 &	-4.45&	-11.44&	-16.29\\
	-2.38&	-0.76&	-1.84&	-2.66\\
	-2.38&	-1.08&	-8.85&	-3.94\\
	-2.38&	-1.34&	9.83&	-0.04 \\
\end{bmatrix} 
$$
The optimal hedge at the last time-step $T$ is $0$, whereas at $t=0$, it is fixed at $-0.05$ due to the constant stock price of $100$. The put option price $C_0$ is equal to the negative of the Q-value, resulting in an option premium of $2.38$.

\citet{dixon2020} highlight the model's tractability, relying on matrix linear algebra, which is indeed evident through the simplified representation of the model-based QLBS.

\subsection{Traditional vs. Reinforcement Learning Option Pricing}

\

This section compares BSM with QLBS prices and hedges.
Following \citeauthor{dixon2020} (\citeyear{dixon2020}), the stock price process is simulated under the probability measure $\mathbb{P}$. Each testing segment considers all three states, as detailed in Section \ref{4.2}.\footnote{For the sake of simplicity, stock price returns will be further denoted as $lnS_t$.}
\begin{table}[H]
	\caption{Input parameters}
	\resizebox{\textwidth}{!}{%
		\begin{tabular}{@{}ccccccccccc@{}}
			\toprule
			\textbf{\begin{tabular}[c]{@{}c@{}}Initial stock \\ price $S_0$\end{tabular}} &
			\textbf{\begin{tabular}[c]{@{}c@{}}Drift of \\ stock $\mu$\end{tabular}} &
			\textbf{\begin{tabular}[c]{@{}c@{}}Volatility of \\ stock $\sigma$\end{tabular}} &
			\textbf{\begin{tabular}[c]{@{}c@{}}Risk-free \\ rate $r$\end{tabular}} &
			\textbf{\begin{tabular}[c]{@{}c@{}}Strike\\ $Z$\end{tabular}} &
			\textbf{Time-steps} &
			\textbf{\begin{tabular}[c]{@{}c@{}}Monte Carlo \\ paths $K$\end{tabular}} &
			\textbf{\begin{tabular}[c]{@{}c@{}}Number of \\ splines $N$\end{tabular}} &
			\textbf{\begin{tabular}[c]{@{}c@{}}Order of \\ splines\end{tabular}} &
			\textbf{\begin{tabular}[c]{@{}c@{}}Noise\\ $\eta$\end{tabular}} \\ \midrule
			100 &
			0.05 &
			0.15 &
			0.03 &
			100 &
			24 &
			10 000 &
			12 &
			4 &
			0.2 \\ \bottomrule
		\end{tabular}%
	}
	\label{tab:experimental}
\end{table}
Table \ref{tab:experimental} lists the selected model parameters used for the numerical implementation, unless otherwise specified. Option maturity is set to 1 year, and the risk aversion parameter $\lambda=0.0001$. Further, in accordance with \citeauthor{halperin1} (\citeyear{halperin1}) and \citeauthor{dixon2020} (\citeyear{dixon2020}), we adjust equation (\ref{a_star}), i.e. (\ref{a_06}), by setting $\frac{1}{2\lambda \gamma}=0$, assuming a pure risk-based hedge.  In the following, we will:
\begin{itemize}
	\item Increase the stock price volatility
	\item Increase the noise for a different number of Monte Carlo paths 
	\item Change the frequency of hedging
	\item Evaluate the performance of QLBS for in- and out-of-the-money options
	\item Incorporate transaction costs
\end{itemize}
Additionally, \ref{app_results} explores the sensitivity of QLBS to changes in the basis functions. 

\

\subsubsection{Volatility and Hedging Frequency}

\

\textbf{Effects of different volatility levels}

\

According to \citeauthor{HullBook} (\citeyear{HullBook}), stock volatility typically ranges from 15\% to 60\%, where higher volatility indicates increased market uncertainty, leading to higher option prices. Therefore, we examine how QLBS performs under different levels of volatility within this range. These tests encompass all three states and involve both the model-free and model-based QLBS. 

Table \ref{tab:vola} compares the QLBS and BSM put option prices at $t=0$ for volatilities $\sigma=0.15$, $\sigma=0.25$, and $\sigma=0.40$.

\begin{table}[H]
	\renewcommand{\arraystretch}{1.1}
	\caption{QLBS vs. BSM put option price at $t=0$, for $\sigma=0.15$, $\sigma=0.25$, and $\sigma=0.40$}
	\label{tab:vola}
	\centering
		\begin{tabular}{@{}lccccccccc@{}}
			\toprule
			\textbf{Volatility} & \multicolumn{3}{c}{$\sigma = 0.15$} & \multicolumn{3}{c}{$\sigma = 0.25$} & \multicolumn{3}{c}{$\sigma =  0.40$} \\ \midrule
			\textbf{States}     & $X_t$     & $S_t$     & $ln S_t$    & $X_t$     & $S_t$     & $ln S_t$    & $X_t$     & $S_t$     & $ln S_t$     \\ \midrule
			Model-Based QLBS    & 4.50      & 4.53      & 4.57        & 8.45      & 8.51      & 8.63        & 14.43     & 14.55     & 14.82        \\
			Model-Free QLBS     & 4.52      & 4.55      & 4.58        & 8.46      & 8.54      & 8.64        & 14.47     & 14.59     & 14.84        \\ \midrule
			BSM       & \multicolumn{3}{c}{4.53}            & \multicolumn{3}{c}{8.39}            & \multicolumn{3}{c}{14.18}            \\ \bottomrule
		\end{tabular}%
  \end{table}
\

As shown in Table \ref{tab:vola}, both the model-based and model-free QLBS option prices are close to the BSM price. For lower levels of $\sigma$, QLBS performs more similarly to the BSM. Furthermore, the model-free QLBS produces marginally higher prices compared to the model-based QLBS. Noticeably, regardless of the chosen state variable, the QLBS price is approximately close to the BSM price. 

\

\newpage
Table \ref{tab:vola_hedge} presents the optimal hedge at $t=0$ for each of the three volatility levels and states.

\begin{table}[H]
	\renewcommand{\arraystretch}{1.1}
	\centering
	\caption{Optimal hedge at $t=0$, for $\sigma=0.15$, $\sigma=0.25$, and $\sigma=0.40$}
	\label{tab:vola_hedge}
		\resizebox{\textwidth}{!}{%
		\begin{tabular}{@{}lccccccccc@{}}
			\toprule
			\textbf{Volatility} & \multicolumn{3}{c}{$\sigma = 0.15$} & \multicolumn{3}{c}{$\sigma = 0.25$} & \multicolumn{3}{c}{$\sigma =  0.40$} \\ \midrule
			\textbf{States}     & $X_t$     & $S_t$     & $ln S_t$    & $X_t$     & $S_t$     & $ln S_t$    & $X_t$     & $S_t$     & $ln S_t$     \\ \midrule
			Model-Based QLBS    & -0.35     & -0.36     & -0.32       & -0.36     & -0.37     & -0.35       & -0.35     & -0.37     & -0.36        \\ \midrule
			BSM       & \multicolumn{3}{c}{-0.39}           & \multicolumn{3}{c}{-0.40}           & \multicolumn{3}{c}{-0.39}            \\ \bottomrule
		\end{tabular}%
			}
\end{table}
Table \ref{tab:vola_hedge} reveals that the state $S_t$ yields a hedge that is slightly closer to the BSM hedge when compared to $X_t$ and $lnS_t$. The reported results pertain only to the model-based QLBS.\footnote{Recall that only the model-based QLBS calculates the optimal hedge $a_t^\star$, whereas the model-free QLBS determines the option price based on the observed hedge.}

\

In summary, the results indicate that both the model-based and model-free QLBS perform well in the presence of increased market uncertainty. The resulting option prices and hedges are closely aligned with the BSM model, with a slightly larger difference in prices for higher volatilities.

\

\textbf{Noise vs. number of Monte Carlo paths}

\

\begin{figure}[H]		\centering	
\hspace{-0.7cm}
\begin{minipage}{0.50\textwidth}
	\centering	\includegraphics[width=7cm,height=7.5cm,keepaspectratio]{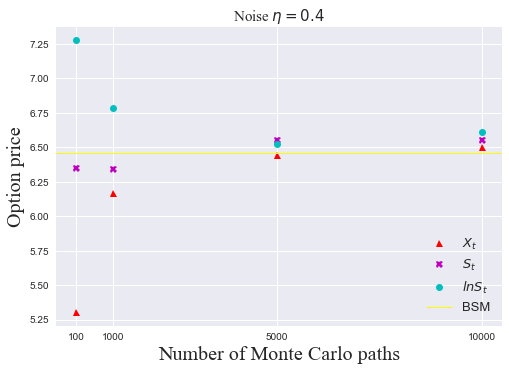}
\end{minipage}
\hfill
\begin{minipage}{0.50\textwidth}
	\centering	\includegraphics[width=7cm,height=7.5cm,keepaspectratio]{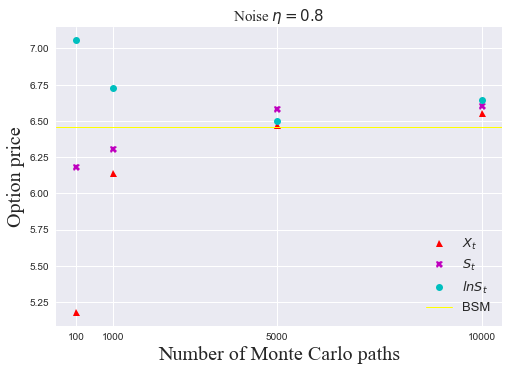}
\end{minipage}
\caption{Model-free QLBS option price for $\sigma=0.2$, with the noise level $\eta = 0.4$ and $\eta=0.8$. The horizontal yellow line represents the continuous-time BSM price}
\label{noise}
\end{figure}

\citeauthor{halperin} (\citeyear{halperin}) states that if enough data are provided, QLBS can learn with purely random actions.
To verify this claim, we conduct tests using $K \in \{100, 1000, 5000, 10000\}$ for the noise levels $\eta \in \{ 0.4, 0.8\}$. The comparison between the model-free QLBS and continuous-time BSM prices is illustrated in Figure \ref{noise}.

Figure \ref{noise} depicts the algorithm's capability to learn the optimal option price, even in the presence of a noise level of 0.8, which indicates nearly random actions (hedges). Furthermore, obtaining a solution comparable to the BSM price does not require $10000$ paths, as it can be attained with as few as $K=5000$.

Overall, these results suggest that QLBS maintains its effectiveness even when confronted with higher noise levels, thereby employing a relatively modest number of Monte Carlo paths.

\

\textbf{Hedging frequency impact on QLBS prices}

\

Additionally, we check the influence of various hedging frequencies on QLBS option prices. The frequencies considered are weekly, bi-weekly, monthly, and semi-annually.

\begin{table}[H]
\caption{QLBS option price at $t=0$ for different hedging frequencies}
\renewcommand{\arraystretch}{1.2}
\label{tab:maturity}
\resizebox{\textwidth}{!}{%
	\begin{tabular}{@{}lcccccccccccc@{}}
		\toprule
		\textbf{Hedging frequency} & \multicolumn{3}{c}{weekly} & \multicolumn{3}{c}{bi-weekly} & \multicolumn{3}{c}{monthly} & \multicolumn{3}{c}{semi-annually} \\ \midrule
		\textbf{States}             & $X_t$  & $S_t$  & $ln S_t$ & $X_t$   & $S_t$   & $ln S_t$  & $X_t$  & $S_t$  & $ln S_t$  & $X_t$     & $S_t$     & $ln S_t$     \\ \midrule
		Model-Based QLBS            & 4.49   & 4.52   & 4.52     & 4.50    & 4.53    & 4.57      & 4.49   & 4.50   & 4.52      & 4.45      & 4.46      & 4.45         \\
		Model-Free QLBS             & 4.48   & 4.51   & 4.51     & 4.52    & 4.55    & 4.58      & 4.50   & 4.51   & 4.53      & 4.45      & 4.46      & 4.45         \\ \bottomrule
	\end{tabular}%
}
\end{table}

Table \ref{tab:maturity} shows that the considered hedging frequencies have a minimal effect on QLBS option pricing, with prices ranging from 4.45 to 4.58, while the BSM price is 4.53.

The observed distinction between the results of model-based and model-free QLBS is negligible. For that reason, the subsequent results will be focused on the model-based QLBS.

\subsubsection{Option's Moneyness}

\

This part explores the effect of different strike prices on the QLBS model for various levels of risk aversion $\lambda$. Specifically, we analyze the QLBS model's behavior for a range of strike prices $Z$, spanning from 60 to 140 with an incremental step of 5. The QLBS model computes both the option prices and hedges for approximately 40 seconds.

\begin{figure}[H]		\centering	
\hspace{-0.7cm}
\begin{minipage}{0.50\textwidth}
	\centering
	\includegraphics[width=7.5cm,height=7.5cm,keepaspectratio]{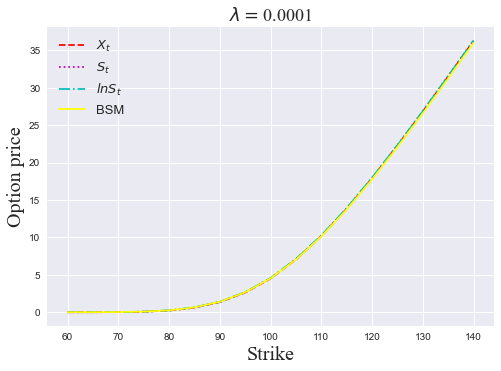}
\end{minipage}
\hfill
\begin{minipage}{0.50\textwidth}
	\centering
	\includegraphics[width=7.5cm,height=7.5cm,keepaspectratio]{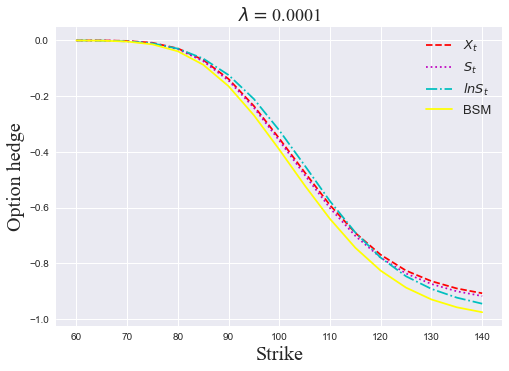}
\end{minipage}
\caption{Model-based QLBS option prices (left) and hedges (right), where $\lambda = 0.0001$ and $S_0=100$}
\label{money1}
\end{figure}
On the left side of Figure \ref{money1} are compared the model-based QLBS prices with the BSM price. For the risk level of $\lambda=0.0001$, the QLBS prices closely align with the BSM prices. QLBS almost perfectly matches the BSM price, confirming indeed \citeauthor{halperin}'s statement (\citeyear{halperin}) that QLBS converges to the BSM price for lower levels of $\lambda$. All three states deliver similar prices.
The QLBS and BSM models' hedge ratios are presented on the right side of Figure \ref{money1}. For deep out-of-the-money options, the QLBS and BSM hedge ratios are nearly identical. However, as the option becomes deeper in-the-money, the hedge ratios start to diverge more noticeably. 

\begin{figure}[H]		\centering
\hspace{-0.7cm}
\begin{minipage}{0.50\textwidth}
	\centering
	\includegraphics[width=7.5cm,height=7.5cm,keepaspectratio]{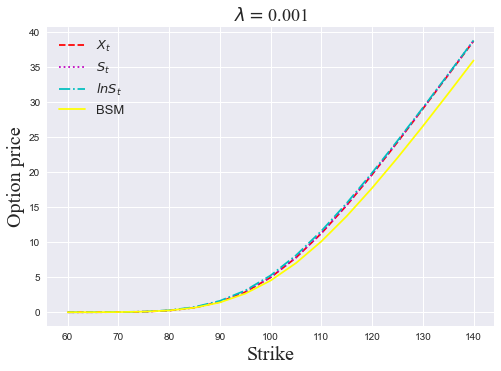}
\end{minipage}
\hfill
\begin{minipage}{0.50\textwidth}
	\centering
	\includegraphics[width=7.5cm,height=7.5cm,keepaspectratio]{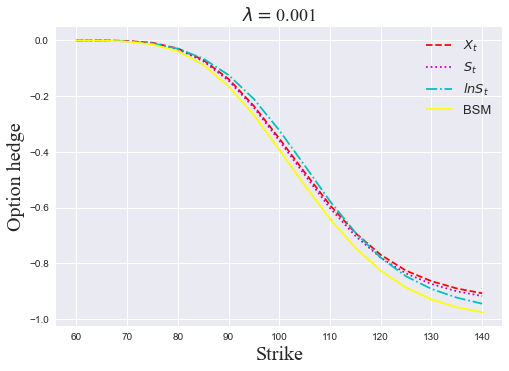}
	
\end{minipage}
\caption{Model-based QLBS option prices (left) and hedges (right), where $\lambda = 0.001$ and $S_0=100$}
\label{money2}
\end{figure}

Figure \ref{money2} illustrates the results for a higher level of $\lambda$. As expected, as risk aversion increases, the QLBS option prices show greater deviation from the BSM prices. This discrepancy is more pronounced for in-the-money options, for which QLBS generates higher prices than BSM. In this light, there is evidence suggesting that under stochastic volatility BSM undervalues deep in-the-money options \citep{hull198}. On the right side of Figure \ref{money2}, the hedges exhibit no differences compared to those in Figure \ref{money1}, which aligns with the assumption of a pure risk-based hedge.

In a nutshell, the QLBS model is capable of pricing and hedging options effectively, regardless of the strike price. Additionally, the results also highlight that increasing the level of risk aversion $\lambda$ has a larger effect on deep in-the-money options.

\

\subsubsection{Transaction Costs}

\

Finally, this part discusses a possible effect of transaction costs on QLBS, which are incorporated after the model-based QLBS optimal actions are calculated.

To determine the terminal wealth ($TW$) of the put option seller, we employ the cash flow approach outlined in \citeauthor{cao2021} (\citeyear{cao2021}):

\begin{equation}
TW_{t} = S_{t}(a_{t-1}^\star-a^\star_{t}) - c |a_{t+1}^\star - a_t^\star|S_{t}
\end{equation}
where $c$ represents a transaction cost. At $t=0$, $TW_0 = -S_0a_0^\star - c |S_0-a_0^\star|$, while at the last time-step T, $TW_T$ is defined as $S_T (a_{T-1}^\star-a_T^\star) - max(Z-S_T;0)$. In addition, the option premium that the writer has received at $t=0$ is added. Here, $c=0.01$ and $\lambda=0.002$ are assumed. 
\begin{table}[H]
\centering
\caption{Mean and median value of the option writer's terminal wealth (in USD) for three different states and $c=0.01$}
\begin{tabular}{@{}llll@{}}
	\toprule
	& $X_t$   & $S_t$   & $lnS_t$ \\ \midrule
	\textbf{mean}   & -0.7942 & -0.6679 & 0.4521  \\
	\textbf{median} & -0.4492 & -0.4169 & 1.5190  \\ \bottomrule
\end{tabular}
\label{tab_PnL}
\end{table}
\begin{figure}[H]
\centering
\includegraphics[width= 7cm, height=7cm, keepaspectratio]{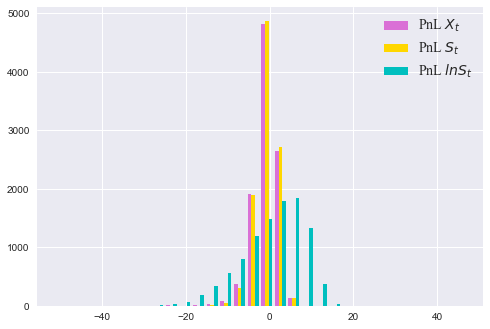}
\caption{Terminal wealth for three different states and $c=0.01$}
\label{fig_pnl}
\end{figure} 

Figure \ref{fig_pnl} highlights a potential importance of choosing the state variable regarding the option writer's $TW$.  
Table \ref{tab_PnL} shows that for the $X_t$ and $S_t$ states, both the median and mean values are negative. However, for $lnS_t$, both values are positive. The higher median value than the mean value suggests a left skewness, which is also evident from Figure \ref{fig_pnl}.

While in the previous analysis the choice of the state variable did not reveal any significant impact, it might be important for $TW$. 

\section{Conclusion}\label{chap_6}

This thesis aims to shed light on the QLBS model by providing the technical notes and its simplified representation. Additionally, the performance of the QLBS model is explored across various scenarios, enabling the synthesis of answers to the research questions defined at the outset:

\

\textit{RQ1)} What are the effects of different levels of \textit{volatility and hedging frequency} on QLBS pricing and hedging?

\begin{itemize}
	\item	The QLBS model accurately \textit{captures the impact of underlying volatility} on option prices.
 \item	The algorithm generates \textit{prices} that \textit{closely align} with those of the \textit{BSM} model, \textit{regardless of} whether \textit{the hedging frequency} is weekly, bi-weekly, monthly, or semi-annually.
	\item	Moreover, the model demonstrates the\textit{ ability to learn from suboptimal actions} when sufficient data is provided.
\end{itemize}

\textit{RQ2)} How does the model perform at  \textit{different moneyness} levels?

\begin{itemize}
\item	The results reveal that QLBS \textit{performs well across various strike prices}, where QLBS pricing differs more from the BSM model for deep in-the-money options when the risk aversion increases.
\end{itemize}

 \textit{RQ3)} What additional effects may arise from incorporating \textit{transaction costs}?

\begin{itemize}
	\item	Finally, it is shown that proportional transaction costs can be considered using the cash flow approach.
	 In this regard, the selection of the state variable could have an impact, as it is demonstrated that employing \textit{$lnSt$ might lead to higher profits} compared to $X_t$ or $S_t$. 
\end{itemize}

\subsection {Further Research}

\

Since this paper compares the performances of QLBS and BSM, constant volatility of the underlying asset is assumed. Therefore, another possibility is to simulate stock prices under stochastic volatility and compare the QLBS approach with the Heston model \citep{heston1993}.

Further work could explore modifications of the original QLBS approach by including transaction costs in the objective function. 

An interesting  extension would be the utilization of Neural Fitted Q-Iteration \citep{riedmiller2005} for function approximation to avoid the dilemma regarding the choice of basis functions.

Lastly, pricing and hedging American-style options within the QLBS framework present promising directions for future research.

\appendix

\section{Abscence of Arbitrage}\label{appendix_BSM}

\

\citeauthor{bjork} (\citeyear{bjork}) argues that according to the\textit{ First Fundamental Theorem}, the absence of arbitrage is possible only under a \textit{martingale measure} $\mathbb{Q}$. Following \citeauthor{bjork}'s arguments,  the Black and Scholes equation in a risk-neutral world can be solved in the style of Feynman-Ka\v{c} solution:  

\begin{equation}
	F(t,s) = exp(-r(T-t)) E^{\mathbb{Q}} \left[ \Phi (S_T) \right]
\end{equation}
with a contingent claim $\Phi (S_T)$.

\

The transition from the real (physical) measure $\mathbb{P}$ to the risk-neutral world $\mathbb{Q}$ in the BSM is possible by employing the Girsanov kernel, which has the interpretation in finance as the\textit{ risk premium per volatility unit}, given by:

\begin{equation}
	\varphi_t = - \frac{\mu - r} {\sigma}
\end{equation}

By applying Girsanov's theorem, the stock price dynamics can be expressed as:

\begin{equation}
	\begin{split}
		dS_t &= \mu S_t dt + \sigma S_t dW_t \quad\quad \quad\quad\quad  \\
		&=  \mu S_t dt + \sigma S_t (\varphi_t dt + dW_t^{\mathbb{Q}}) \quad\quad\\
		&= \mu S_t dt + \sigma S_t \varphi_t dt + \sigma S_t dW_t^{\mathbb{Q}} \\
		&= S_t (\mu + \sigma \varphi_t) dt + \sigma S_t dW_t^{\mathbb{Q}} \\
	\end{split}
\end{equation}

This leads to the stochastic differential equation for $S_t$ in the risk-neutral world:
\begin{equation}
	\begin{split}
		dS_t &= S_t r dt + \sigma S_t dW_t^{\mathbb{Q}} 
	\end{split}
\end{equation}
where $dW_t^{\mathbb{Q}}$ is a Wiener process under the probability measure $\mathbb{Q}$ (or a Q-Wiener process).

The QLBS model is not assumed to be risk neutral, and therefore, the dynamics of the stock price are modeled under  the real probability measure  $\mathbb{P}$.

\section{Technical Notes}\label{app_technical}
\subsection{Defining Replicating Portfolio}\label{app_1}

\

The replicating portfolio $\Pi_t$ is defined in equation  (\ref{eq01}), and we assume that the portfolio at $t+1$ is $\Pi_{t+1} = a_{t+1}S_{t+1} + B_{t+1}$. The initial bank account $B_t$ is risk-free and it is expected to earn a risk-free rate, so that $B_{t+1} =exp(r\Delta t) B_t $. By applying the self-financing constraint, as in \citet{halperin}:
$$a_tS_{t+1} + B_t exp(r\Delta t) = a_{t+1}S_{t+1}+B_{t+1}$$ 
follows that:
\begin{equation}
	\label{cash}
	B_t = \gamma (B_{t+1} + (a_{t+1} S_{t+1}- a_{t}S_{t+1})) 
\end{equation}
By plugging equation (\ref{cash}) into  (\ref{eq01}), we obtain:
\begin{equation} 
	\begin{split}
		\Pi_t & = a_t S_t + \gamma (B_{t+1} + a_{t+1} S_{t+1} - a_t S_{t+1}) \quad\quad \text{$\triangleright$ Apply $\Pi_{t+1} = a_{t+1}S_{t+1} + B_{t+1} $} \\
		& = \gamma (\Pi_{t+1} -  a_t S_{t+1} + a_t S_t exp(r \Delta t))\\
		& = \gamma (\Pi_{t+1} - a_t(S_{t+1} - S_t exp(r\Delta t))) \\
		& = \gamma (\Pi_{t+1} - a_t \Delta S_t)
	\end{split}
	\label{pi}
\end{equation}
In this way, equation (\ref{eq_pi}) is recovered.

\

\subsection{Action-Value Function with a Shifted Time Argument} \label{append_action_value}

\

From equation (\ref{qlbs_q_value}), we can express the action-value function at $t+1$ as:
\begin{equation}
	\begin{split}
		Q^\pi_{t+1} (X_{t+1}, a_{t+1}) = \mathbb{E}_{t+1} \left[ -\Pi_{t+1} (X_{t+1}) - \lambda \sum_{t'=t}^T exp(-r(t'-(t+1))) Var [\Pi_{t'}(X_{t'}) |\mathcal{F}_{t'}] | \mathcal{F}_t \right] 	
	\end{split}
\end{equation}
so that:
\begin{equation}
	\begin{split}
		& Q^\pi_{t+1} (X_{t+1}, a_{t+1}) + \mathbb{E}_{t+1} \left[\Pi_{t+1}(X_{t+1}) \right] = - \lambda \mathbb{E}_{t+1} \left[ \sum_{t'=t}^T exp(-r(t'-(t+1))) Var [\Pi_{t'}(X_{t'}) |\mathcal{F}_{t'}] | \mathcal{F}_t \right]  \\
	\end{split}
\end{equation}
and consequently:
\begin{equation}
	\gamma \left(Q^\pi_{t+1} (X_{t+1}, a_{t+1}) + \mathbb{E}_{t+1} \left[\Pi_{t+1}(X_{t+1}) \right] \right) = -\lambda \mathbb{E}_{t+1} \left[ \sum_{t'=t+1}^T exp(-r(t'-t)) Var [\Pi_{t'}(X_{t'}) |\mathcal{F}_{t'}] \right] 
	\label{eq_next_q_value}
\end{equation}
which is then applied in equation (\ref{qlbs_q_value}).

\

\subsection{Defining Reward} \label{app_reward}

\

From the reward definition given in equation (\ref{reward}), we have:
$$R_t\left(X_t,a_t,X_{t+1}\right)=\gamma a_t\Delta S_t (X_t, X_{t+1})-\lambda Var\left[\Pi_t\space|\space\mathcal F_t\right], \quad \quad t=T-1,...,0$$

To determine the variance of the portfolio $\Pi_t$, we employ $ Var(X) = \mathbb{E}((X-\mathbb{E} (X))^2) $ and $Var(X-bY) = Var(X) - 2bCov(X,Y) + b^2Var(Y)$, and thus we obtain: 

\begin{equation}
		\resizebox{1.1\textwidth}{!}{%
		$
	\begin{split}
		Var(\Pi_t | \mathcal{F}_t) & = Var [(\Pi_{t+1}-a_t \Delta S_t) | \mathcal{F}_t] \\
		& = [Var_t[\Pi_{t+1}] - 2a_tCov_t[\Pi_{t+1}, \Delta S_t] + a_t^2Var_t[\Delta S_t] ] \\
		& = \mathbb{E}_t  \left[\left(\Pi_{t+1} - \mathbb{E}_t(\Pi_{t+1})\right)^2 - 2a_t\left( (\Pi_{t+1} - \mathbb{E}_t(\Pi_{t+1})) (\Delta S_t - \mathbb{E}_t(\Delta S_t))\right) + a_t^2\left(\Delta S_t - \mathbb{E}_t(\Delta S_t)\right)^2 \right] \\
		& = \mathbb{E}_t [(\Pi_{t+1} - \overline{\Pi}_{t+1})^2 - 2 a_t (\Pi_{t+1} - \overline{\Pi}_{t+1})(\Delta S_t - \Delta \overline{S_t}) + a_t^2 (\Delta S_t -\Delta \overline{S_t} )^2   ] \\
		& = \mathbb{E}_t [\hat{\Pi}_{t+1}^2 -2 a_t\hat{\Pi}_{t+1} \Delta \hat{S}_t + a_t^2(\Delta \hat{S}_t)^2]	
	\end{split}
$%
}
	\label{eq_variance_portfolio}
\end{equation}
where $\mathbb{E} (\Pi_{t+1}) =\overline{\Pi}_{t+1}$ represents the mean value of the portfolio $\Pi$ at $t=t+1$, and  $\mathbb{E}(\Delta S_t) = \Delta \overline{S}_{t}$ is the mean of $\Delta S$ across all Monte Carlos paths at $t$. The terminal reward is $R_T=-\lambda Var\left[\Pi_T\right]$.

\

\subsection{Optimal Hedge} \label{app_optimal_hedge}

\

The optimal hedge is obtained by setting the partial derivative of equation (\ref{eq_q_value}) with respect to $a_t$ to zero. Firstly, we will rearrange (\ref{eq_q_value}) as follows:

\begin{equation}
\resizebox{1.05\textwidth}{!}{%
	$
	\begin{split}
		\label{eq_q_01}
		Q_t^\star (X_t, a_t)
		& = \gamma \mathbb{E}_{t} \left[ 	Q_{t+1}^\star (X_{t+1}, a_{t+1}^\star) + a_t \Delta S_t\right] -\mathbb{E}_t\left[ \lambda \gamma^2\hat{\Pi}^2_{t+1} - \lambda \gamma^2 2a_t\hat{\Pi}_{t+1} \Delta \hat{S}_t + \lambda \gamma^2 a_t^2 (\Delta \hat{S}_t)^2 \right] \\
	\end{split}
$ }
\end{equation}

Further, by setting $\frac{\partial \left(- Q^\star_t (X_t, a_t)\right)}{\partial a_t} = 0 $, we have:
\begin{equation}
	\label{technical_01}
	\begin{aligned}
		\frac{\partial \left(- Q^\star_t (X_t, a_t)\right)}{\partial a_t} &=	\mathbb{E}_{t} \left[-\gamma \Delta S_t - 2\lambda \gamma^2  \Delta \hat{ S}_t\hat{\Pi}_{t+1} + 2\lambda \gamma^2 a_t (\Delta \hat{ S}_t)^2 \right] = 0 \\
	\end{aligned}
\end{equation}
and then:
\begin{equation}
	\begin{aligned}
		& \mathbb{E}_t\left[2\lambda \gamma^2 a_t (\Delta \hat{S}_t)^2\right] =  \mathbb{E}_t \left[\gamma \Delta S_t + 2\lambda \gamma^2  \Delta \hat{S}_t\hat{\Pi}_{t+1} \right] \\	
	\end{aligned}
\end{equation}

\

Next, by rearranging the terms, we obtain:
\begin{equation}
	\setstretch{3}
	\begin {split}
	a_t ^\star (X_t) &= \frac{\mathbb{E}_t \left[\gamma \: \Delta S_t \: + \: 2\lambda \gamma^2\:\Delta \hat{S}_t\: \hat{\Pi}_{t+1} \right]}{2\:\lambda \,\gamma^2\:\mathbb{E}_t \:\left[(\Delta \hat{S}_t)^2\right]}\\
	&=\frac{\gamma\:\mathbb{E}_t \left[\Delta S_t + 2 \, \lambda\,\gamma \,\Delta\hat{S}_t\,\hat{\Pi}_{t+1} \right] }{2\,\lambda\, \gamma^2\,\mathbb{E}_t\,\left[(\Delta \hat{ S}_t)^2\right]} \\
	&=\frac{\mathbb{E}_t \,\left[\Delta S_t + \, 2 \,\lambda\,\gamma \,\Delta\hat{S}_t\, \hat{\Pi}_{t+1} \right] }{2\,\lambda\,\gamma\,\mathbb{E}_t\,\left[(\Delta\, \hat{ S}_t)^2 \right]} \\
	&= \frac{\mathbb{E}_t\,\left[\,\frac{1}{2\,\lambda\,\gamma}\,\Delta S_t + 2 \frac{1}{2\,\lambda\,\gamma}\, \lambda\, \gamma\, \Delta \hat{S}_t\, \hat{\Pi}_{t+1}\right]}{\mathbb{E}_t\,\left[(\Delta \,\hat{ S}_t)^2\right]}\\
	&= \frac{\mathbb{E}_t\, \left[\frac{1}{2\,\lambda\,\gamma}\,\Delta S_t + \,\Delta \hat{S}_t\,\hat{\Pi}_{t+1} \right]}{\mathbb{E}_t\,\left[(\Delta \hat{ S}_t)^2 \right]}
\end{split}
\end{equation} 

Finally, the analytical solution for $a^\star_t$ is achieved, as given in equation (\ref{a_star}). 

\

\subsection{Solution for the Optimal Action }\label{app_coeff_phi}

\

As discussed in Chapter \ref{4.2}, Monte Carlo simulation is used to obtain the optimal actions in practice. By changing the sign and substituting $\mathbb{E}_t (\cdotB)$ with Monte Carlo simulation, equation (\ref{eq_q_value}) becomes:

\begin{equation}
\label{q_03}
\resizebox{1.05\textwidth}{!}{%
	$
\begin{split}
	- 	Q_t^\star (X_t, a_t) 	& =\mathbb{E}_{t} \left[ - \gamma 	Q_{t+1}^\star (X_{t+1}, a_{t+1}^\star) -  \gamma a_t \Delta S_t\right] + \lambda \gamma^2 \mathbb{E}_t \left[ \left(  \hat{\Pi}_{t+1} - a_t (\Delta \hat{S}_t)\right)^2 \right] \quad \text{$\triangleright$ Apply (\ref{basis_01}) for $a_t$ } \\
	&= \sum_{k=1}^{K} \left[ -\gamma  	Q_{t+1}^{\star} (X_{t+1}, a_{t+1}^{\star}) - \gamma \sum_n{\phi_{nt}\Phi_n\left(X_t^k\right)}\Delta S_t^k + \lambda \gamma^2  \left(  \hat{\Pi}_{t+1}^k - \sum_n {\phi_{nt}\Phi_n\left(X_t^k\right)} (\Delta \hat{S}_t^k)\right)^2  \right] \\
\end{split}
$%
}
\end{equation}

\

Considering only the action-dependent terms, it follows that:

\begin{equation}
\label{q_04}
\begin{split}
	G_t (\phi_{nt}) &= \sum_{k=1}^{K} \left[ - \sum_n {\phi_{nt}\Phi_n\left(X_t^k\right)}\Delta S_t^k  + \lambda \gamma  \left(  \hat{\Pi}_{t+1}^k - \sum_n {\phi_{nt}\Phi_n\left(X_t^k\right)} (\Delta \hat{S}_t^k)\right)^2\right]   \\		
	& = \sum_{k=1}^{K} \Bigg[ - \sum_n {\phi_{nt}\Phi_n\left(X_t^k\right)}\Delta S_t^k  + \lambda \gamma   \hat{\Pi}_{t+1}^{2,k} - 2 \lambda \gamma \hat{\Pi}_{t+1}^k \sum_n  {\phi_{nt}\Phi_n \left(X_t^k\right)} \left(\Delta \hat{S}_t^k\right) \\
	&  \quad \quad \quad + \lambda \gamma  \sum_n {\phi_{nt}^2\Phi_n \left(X_t^k\right) \Phi_m \left(X_t^k\right) } \left(\Delta \hat{S}_t^k\right)^2  \Bigg]   \\
\end{split}
\end{equation}
Next, set the derivative of (\ref{q_04}) with respect to $\phi_{nt}$ to zero:
\begin{equation}
\label{q_05}
\begin{split}
	\frac{\partial G_t (\phi_{nt})}{\partial \phi_{nt} } = \sum_{k=1}^{K} & \Bigg[ - \sum_n {\Phi_n\left(X_t^k\right)}\Delta S_t^k  - 2 \lambda \gamma   \hat{\Pi}_{t+1}^k \sum_n  {\Phi_n \left(X_t^k\right)} \left(\Delta \hat{S}_t^k\right)\\
	& + 2 \lambda \gamma 	\sum_n {\phi_{nt}\Phi_n \left(X_t^k\right) \Phi_m \left(X_t^k\right) } \left(\Delta \hat{S}_t^k\right)^2  \Bigg] = 0    \\		
\end{split}
\end{equation}
Finally, the parameters $\phi_{t}^\star$ are obtained as follows:
\begin{equation}
\label{a_06}
\begin{split}
	\phi_t^\star & = \Bigg[ \sum_{k=1}^{K}    2 \lambda \gamma 	{\Phi_n \left(X_t^k\right) \Phi_m \left(X_t^k\right) } \left(\Delta \hat{S}_t^k\right)^2    \Bigg]^{-1} \\
	& \quad  \Bigg[ \sum_{k=1}^{K}     {\Phi_n\left(X_t^k\right)}\Delta S_t^k  + 2 \lambda \gamma   \hat{\Pi}_{t+1}^k  {\Phi_n \left(X_t^k\right)} \left(\Delta \hat{S}_t^k\right) \Bigg] \\   
	&= \Bigg[   \sum_{k=1}^{K}   {\Phi_n \left(X_t^k\right) \Phi_m \left(X_t^k\right) } \left(\Delta \hat{S}_t^k\right)^2    \Bigg]^{-1}  \\		
	&	\quad \Bigg[ \sum_{k=1}^{K}  \frac{1}{2\lambda \gamma}    {\Phi_n\left(X_t^k\right)}\Delta S_t^k  + 2 \frac{1}{2\lambda \gamma} \lambda \gamma   \hat{\Pi}_{t+1}^k   {\Phi_n \left(X_t^k\right)} \left(\Delta \hat{S}_t^k\right) \Bigg] \\ 
	& =  \Bigg[ \sum_{k=1}^{K}   	 {\Phi_n \left(X_t^k\right) \Phi_m \left(X_t^k\right) } \left(\Delta \hat{S}_t^k\right)^2    \Bigg]^{-1} \\
	& \quad \Bigg[ \sum_{k=1}^{K}   {\Phi_n\left(X_t^k\right)} \Big[ \frac{1}{2\lambda \gamma}   \Delta S_t^k  + \hat{\Pi}_{t+1}^k  \left(\Delta \hat{S}_t^k\right) \Big] \Bigg]  \\
\end{split}
\end{equation}
with $\Phi_m\left(X_t^k\right) = \big[\Phi_n\left(X_t^k\right) \big]^T $, and vice versa. 

\

\subsection{Optimal Coefficients for Action-Value Function}\label{app_optimal_omega}

\

As described in \citeauthor{dixon2020} (\citeyear{dixon2020}), the optimal action-value function can be interpreted as regression:
\begin{equation}
\begin{split}
	R_t(X_t,a_t^\star, X_{t+1}) + \gamma \max_{a_{t+1} \in \mathcal{A}} Q_{t+1}^\star (X_{t+1}, a_{t+1}) &= Q_t^\star (X_t, a_t^\star) + \epsilon_t \\
	&= \sum_n^N{\omega_{nt}\Phi_n\left(X_t\right)} + \epsilon_t \\
\end{split}
\label{eq_q_reg}
\end{equation}
with $\mathbb{E}(\epsilon_t) = 0$. To find the coefficients $\omega_{nt}$, we need to minimize the loss function for each time-step:
\begin{equation}
F_t(\omega) = \sum_{k=1}^{K} \left( R_t(X_t,a_t^\star, X_{t+1}) + \gamma \max_{a_{t+1} \in \mathcal{A}} Q_{t+1}^\star (X_{t+1}, a_{t+1}) -  \sum_n^N{\omega_{nt}\Phi_n\left(X_t^k\right)} \right)^2
\end{equation}
We can simplify this expression so that $\textbf{Y}_t=  \sum_{k=1}^{K} \left( R_t(X_t,a_t^\star, X_{t+1}) + \gamma \max_{a_{t+1} \in \mathcal{A}} Q_{t+1}^\star (X_{t+1}, a_{t+1}) \right) $, with the transposed features $\textbf{P}_t^T = \sum_{k=1}^{K} \Phi_n\left(X_t^k\right)$. Here, we use the well-known least squares optimization:
\begin{equation}
\omega_t^\star = \big(\textbf{P}_t^T \textbf{P}_t \big)^{-1} \textbf{P}_t^T \textbf{Y}_t
\label{app_eq_omega}
\end{equation}
where $\textbf{P}_t^T$ can be represented in matrix form as:\footnote{To be able to invert $\big(\textbf{P}_t^T \textbf{P}_t \big)^{-1}$, we need to add an identity matrix with a small regularization parameter.}
\begin{equation}
\textbf{P}_t^T = \begin{bmatrix}
	\Phi_1\left(X_t^1\right) & \cdots &\Phi_1\left(X_t^K\right)\\
	\Phi_2\left(X_t^1\right) & \cdots &\Phi_2\left(X_t^K\right)\\
	\vdots & &\vdots \\
	\Phi_N\left(X_t^1\right) &  \cdots &\Phi_N\left(X_t^K\right)\\
\end{bmatrix}
\end{equation}
with dimensions $N \times K$. 

The obtained vector of time-dependent coefficients $\omega_t^\star$ is:
\begin{equation}
\omega_t^\star = \begin{bmatrix}
	\omega_1 \\
	\omega_2 \\
	\vdots \\
	\omega_N \\
\end{bmatrix}
\end{equation}
with dimensions $N \times 1$.

\section{Changes in Basis Functions}\label{app_results}

This part explores the sensitivity of QLBS prices and hedges to variations in the order and number of splines.\footnote{ Note that increasing the order and number of splines induces higher computational costs.} The selected parameters include $N=\{15,20,50,100\}$ for the orders $\{1,3,10\}$.

\begin{figure}[H]		\centering	
\hspace{-0.7cm}
\begin{minipage}{0.50\textwidth}
	\centering
	\includegraphics[width=7.5cm,height=7.5cm,keepaspectratio]{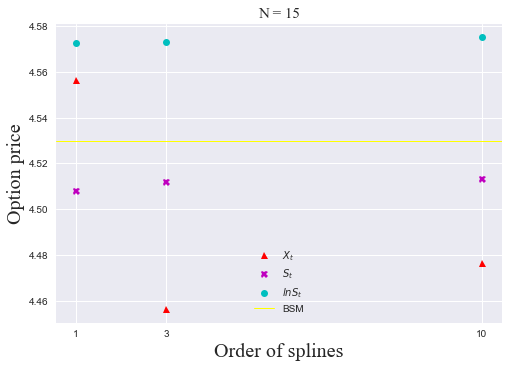}
\end{minipage}
\hfill
\begin{minipage}{0.50\textwidth}
	\centering
	\includegraphics[width=7.5cm,height=7.5cm,keepaspectratio]{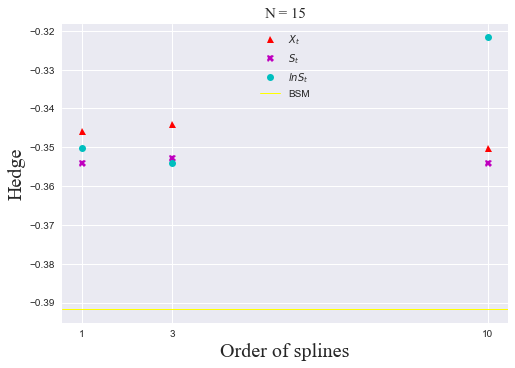}
\end{minipage}
\caption{Model-based QLBS option price and hedge at $t=0$ with $N=15$ basis functions}
\label{basis1}
\end{figure}
Figure \ref{basis1} displays the model-based QLBS option prices and hedges at $t=0$ with $N=15$ basis functions. The left side of the figure shows the QLBS prices, where the horizontal yellow line represents the BSM price. It can be observed that the state $X_t$ is more sensitive to a higher order of splines compared to $S_t$ and $lnS_t$. On the right side of Figure \ref{basis1} are presented the QLBS and BSM hedges at $t=0$. Regardless of the spline order, the BSM hedge consistently has a more negative value compared to the QLBS hedges.

\begin{figure}[H]		\centering	
\hspace{-0.7cm}
\begin{minipage}{0.50\textwidth}
	\centering
	\includegraphics[width=7.5cm,height=7.5cm,keepaspectratio]{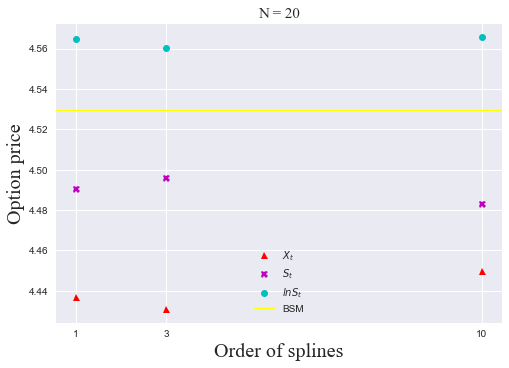}
\end{minipage}
\hfill
\begin{minipage}{0.50\textwidth}
	\centering	\includegraphics[width=7.5cm,height=7.5cm,keepaspectratio]{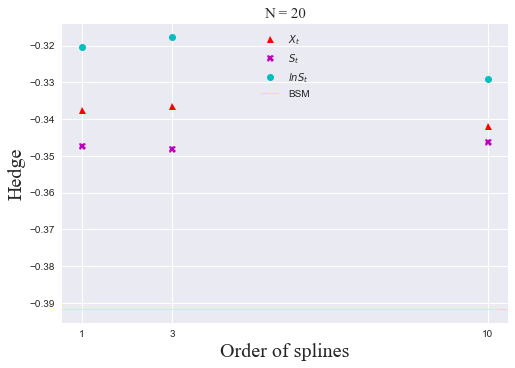}
\end{minipage}
\caption{Model-based QLBS option price and hedge at $t=0$ with $N=20$ basis functions}
\label{basis2}
\end{figure}

Figure \ref{basis2} illustrates that with $N=20$, the deviation of the QLBS prices from the BSM price is slightly more pronounced compared to $N=15$, except for $lnS_t$. Similarly, the QLBS hedge exhibits a marginally higher difference from the BSM hedge compared to Figure \ref{basis1}.

\begin{figure}[H]		\centering	
\hspace{-0.7cm}
\begin{minipage}{0.50\textwidth}
	\centering
	\includegraphics[width=7.5cm,height=7.5cm,keepaspectratio]{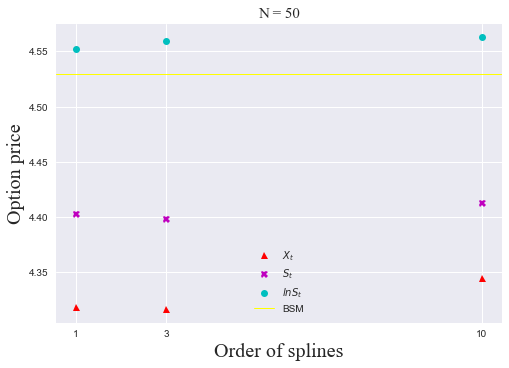}
\end{minipage}
\hfill
\begin{minipage}{0.50\textwidth}
	\centering
	\includegraphics[width=7.5cm,height=7.5cm,keepaspectratio]{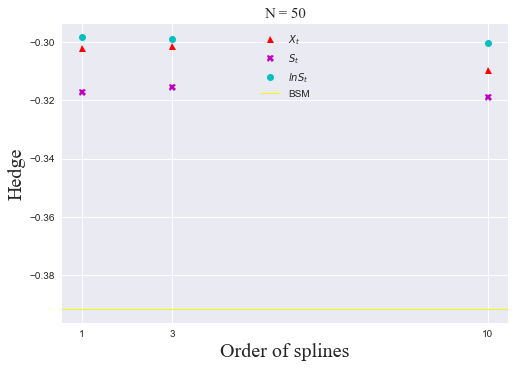}
\end{minipage}
\caption{Model-based QLBS option price and hedge at $t=0$ with $N=50$ basis functions}
\label{basis3}
\end{figure}

Figure \ref{basis3} presents the results for $N=50$ basis functions. Increasing the number of basis functions appears to have a greater impact on $X_t$ and $S_t$ than on $lnS_t$. As in the previous case, $S_t$ demonstrates the hedge closest to the BSM hedge. 

\begin{figure}[H]		\centering	
\hspace{-0.7cm}
\begin{minipage}{0.50\textwidth}
	\centering
	\includegraphics[width=7.5cm,height=7.5cm,keepaspectratio]{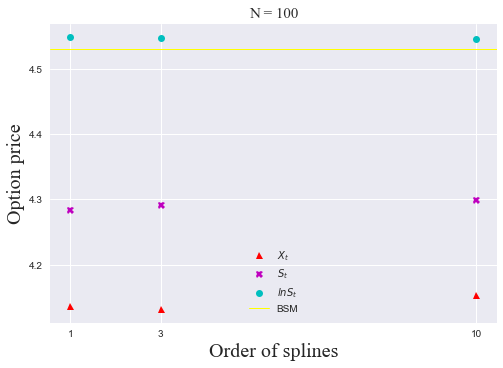}
\end{minipage}
\hfill
\begin{minipage}{0.50\textwidth}
	\centering
	\includegraphics[width=7.5cm,height=7.5cm,keepaspectratio]{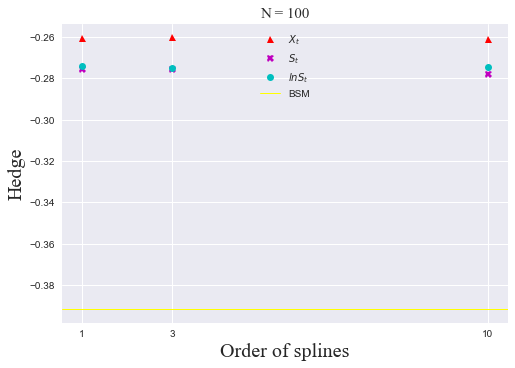}
\end{minipage}
\caption{Model-based QLBS option price and hedge at $t=0$ with $N=100$ basis functions}
\label{basis4}
\end{figure}
Finally, Figure \ref{basis4} reaffirms the previously observed impact of increasing the number of splines for each state variable. With $N=100$, the deviation from the BSM price increased even further for the states $X_t$ and $S_t$, while $lnS_t$ remains almost unchanged. Furthermore, increasing the number of splines causes the hedges to deviate more from the BSM hedge.

In summary, for $N=15$, altering the order of splines can impact QLBS pricing for $X_t$ and affect QLBS hedging for $lnS_t$. However, for a higher number of basis functions, the spline order does not show a significant influence. As the number of splines increases, $lnS_t$ yields prices closest to the BSM price.

\bibliography{Literatur_arXiv}	

\bibliographystyle{elsarticle-harv}\biboptions{authoryear}

\end{document}